\documentclass{article}
\usepackage[utf8]{inputenc}
\usepackage[mathscr]{eucal}
\usepackage[all]{xy}
\usepackage[dvipsnames]{xcolor}
\usepackage{amsmath, amsthm, amssymb, amscd, amsxtra,color,authblk,tikz,arydshln,array, caption, romannum,verbatim,enumitem}
\usepackage{graphicx}
\usepackage{subcaption}
\usetikzlibrary{shapes,arrows,positioning}
\usepackage[T1]{fontenc}
\usepackage{lmodern}
\usepackage{comment}

\usepackage[section]{placeins}
\usetikzlibrary{decorations.markings}
\usepackage{appendix}
\usepackage{booktabs}

\usepackage[sorting=none]{biblatex}
\title{Generalized Kitaev Spin Liquid model and Emergent Twist Defect}

\author[1]{Bowen Yan $ ^*$}
\author[1]{Penghua Chen}
\author[1,2]{Shawn X. Cui \thanks{Corresponding author}}
\affil[1]
{{\small Department of Physics and Astronomy, Purdue University, West Lafayette}}
\affil[2]{{\small Department of Mathematics, Purdue University, West Lafayette}}
\affil[ ]{{\small \it \{yan312, chen3014, cui177\} @purdue.edu}}
\date{}
\begin{document}
\pagenumbering{arabic}

\maketitle
\begin{abstract}
The Kitaev spin liquid model on honeycomb lattice offers an intriguing feature that encapsulates both Abelian and non-Abelian phases\cite{Kitaev_2006}. Recent studies suggest that the comprehensive phase diagram of possible generalized Kitaev model largely depends on the specific details of the discrete lattice, which somewhat deviates from the traditional understanding of ``topological'' phases. In this paper, we propose an adapted version of the Kitaev spin liquid model on arbitrary planar lattices. Our revised model recovers the surface code model under certain parameter selections within the Hamiltonian terms.  Changes in parameters can initiate the emergence of holes, domain walls, or twist defects. Notably, the twist defect, which presents as a lattice dislocation defect, exhibits non-Abelian braiding statistics upon tuning the coefficients of the Hamiltonian. Additionally, we illustrate that the creation, movement, and fusion of these defects can be accomplished through natural time evolution by linearly interpolating the static Hamiltonian. These defects demonstrate the Ising anyon fusion rule as anticipated. Our findings hint at possible implementation in actual physical materials owing to a more realistically achievable two-body interaction.
\end{abstract}

\section{Introduction}
Since Kitaev proposed the Kitaev Quantum Double  model\cite{Kitaev_2003}, it has garnered considerable attention due to its typical anyon behavior and the paradigm it provides for topological quantum computation. The model demonstrates how one can circumvent local errors by encoding information into anyon types and executing gates through anyon braiding, whose information is completely described by Unitary Modular Tensor Categories (UMTC). It has been proven that  certain non-Abelian cases, such as Fibonacci Anyon,  can support universal quantum computation.

Following this development, numerous lattice models have been proposed with the objective of identifying different types of anyons. Two significant classes of these include the Kitaev Quantum Double model\cite{Kitaev_2003} and the Levin-Wen model\cite{Levin_Wen_2005}. These models actualize anyon models from varying perspectives, which are described by the Drinfeld center of a fusion category.

%Realizing the actual topological phase is a challenging and significant mission. Those well known model such as Kitaev Quantum Double model and Levin-Wen model require multi-body interactions and are hard to be settle down in real laboratory. toric code is the simplest but far away from universal computation since it supports abelian anyons.  That is why the twist defect which is introduced by \cite{Bombin_2010} arises people's attention. It manifest non-abelian Ising anyon  and originated from the lattice dislocation of the toric code model and it has been recently observed the Ising anyon statistics experimentally in \cite{andersen_non-abelian_2023}. We should notice, the defect resides on the breaking of two-colorability of the lattice. 

The realization of the actual topological phase is a complex and pivotal task. Renowned models, such as the Kitaev Quantum Double model and Levin-Wen model, necessitate multi-body interactions, making them challenging to implement in a real-world laboratory setting. While some comparatively achievable cases, such as the toric code, are not suitable for universal computation because it only supports Abelian anyons. This reality has led to an increased interest in the twist defect, as introduced by \cite{Bombin_2010} . This defect exemplifies a non-Abelian Ising anyon, which stems from the lattice dislocation of, the Abelian anyon case, the toric code model. Recent experimental observations of the Ising anyon statistics, as reported by \cite{andersen_non-abelian_2023}, attest to this. It should be noted that the defect is dependent on the disruption of the lattice's local two-colorability.

Another intriguing model is the Kitaev spin liquid Model\cite{Kitaev_2006}, which supports Abelian anyons in the gapped phase region, as well as non-Abelian anyons upon the introduction of a magnetic field to the gapless phase. This model is simple yet fruitful. But the definition of the model relies heavily on the geometry of honeycomb lattice, which deviates the idea of topological phase and is the main question to be solved in this paper. Moreover, it also has been pointed out that a spin liquid model on honeycomb lattice with lattice dislocation will generate the twist defect as in \cite{Zheng_Dua_Jiang_2015}. The generalization to Zetor model has been shown in \cite{You_Wen_2012}. This model is potentially easily realizable in a real laboratory due to the two-body nearest interaction.

Considerable theoretical progress has been made in the generalization of this model. Examples include those on a translationally invariant two dimensional lattice with higher-coordination vertices \cite{Chua_Yao_Fiete_2011}\cite{Wu_Arovas_Hung_2009}\cite{Yao_Zhang_Kivelson_2009a}, on a two-dimensional amorphous lattice \cite{Cassella_D’Ornellas_Hodson_Natori_Knolle_2022}, a three-dimensional diamond lattice \cite{Ryu_2009}, and works on trivalent 3D lattices \cite{Eschmann_Mishchenko_O’Brien_Bojesen_Kato_Hermanns_Motome_Trebst_2020}. It is clear that the overall phase diagram is strongly influenced by the geometric specifics of the lattice, thus also deviating our traditional understanding of `topological' phases.

In this paper, we demonstrate that the entire theory can be formulated on a generic planar lattice. The main motivation relies on the toric code limit of the original honeycomb spin liquid model as mentioned before, which is briefly reviewed in section(\ref{sec:honeycomb model}). We sketch the main idea here and details are in the following sections.

The Hamiltonian of the honeycomb spin liquid system is a summation of weighted check operators, which are two-body nearest Pauli operators. The Hamiltonian is frustrated due to non-commutation of the check operators. We say a check operator in the Hamiltonian is dominant if the coefficient of the operator is much larger than others.  Kitaev selected what he refers to as ``z-link'' check operators to take dominance in the Hamiltonian. As a result, the vicinity of the ground state in the spectrum can be accurately described by a toric code model. The exact choice of ``z-link'' check operators is not important. The key is that ``z-link'' check operators composite a maximum set of commuting operators, which is denoted as $\mathbf{stabilizer\, center}$ $S_c$ in this paper. $S_c$ satisfies that any check operator outside this set should anticommutes with exactly two elements in $S_c$. We find that if one can find a proper $S_c$ on an arbitrary planar lattice, a toric code model always appears in the vicinity of the ground state, provided all elements in the $S_c$ are dominant. 

Moreover, we get toric code with defects if we slight break the requirement of $S_c$. Further, we propose that a linear interpolating Hamiltonians, which statically has different dominant $S_c$s, could be a natural way to create, move, and fuse defects in a physical system. This approach circumvents the need for geometric deformation or the application of a coding method. This proposal might inspire real material realization since we only need to establish and adjust the strength of two-body interactions, as illustrated in section \ref{sec:twist defect}. Moreover, a circuit description is plausible since these operations are facilitated by time evolution operators, which are naturally unitary.

This paper is organized as follows:

In Section \ref{sec:honeycomb model}, we provide a concise review of the original honeycomb model and reintroduce  necessary notations, such as the shrunken lattice.

Section \ref{sec:generalization} introduces our method of generalization. Initially, we rewrite the toric code on a lattice where qubits are positioned on vertices instead of edges, as discussed in Section \ref{sec: rewrite the toric code}. This rewriting is inspiring since the recovery of toric code typically ends in a lattice where qubits are placed on vertices. Subsequently, in Sections \ref{sec:even codimension} and \ref{sec:on general lattice}, we demonstrate how to define check operators on arbitrary lattices and, given an appropriate choice of a Stabilizer Center ($S_c$), the toric code can be recovered when the shrunken lattice is 2-colorable. We should note that a local disruption of the two-colorability of the shrunken lattice leads to the emergence of a twist defect.

Section \ref{sec:twist defect} illustrates the process of creation, movement, and fusion of defects through time evolution operators.

Next, in Section \ref{sec:subsystem code}, we demonstrate that the entire model can be treated as a zero-logic-qubit subsystem code in the context of error correcting code.

Finally, Section \ref{sec:conclusion and outlook} concludes the paper and discusses potential future extensions.

\section{Kitaev Honeycomb Model}
\label{sec:honeycomb model}

\begin{figure}[htbp]
    \centering
    \begin{subfigure}{0.3\textwidth}
        \centering
        \includegraphics[width=\linewidth]{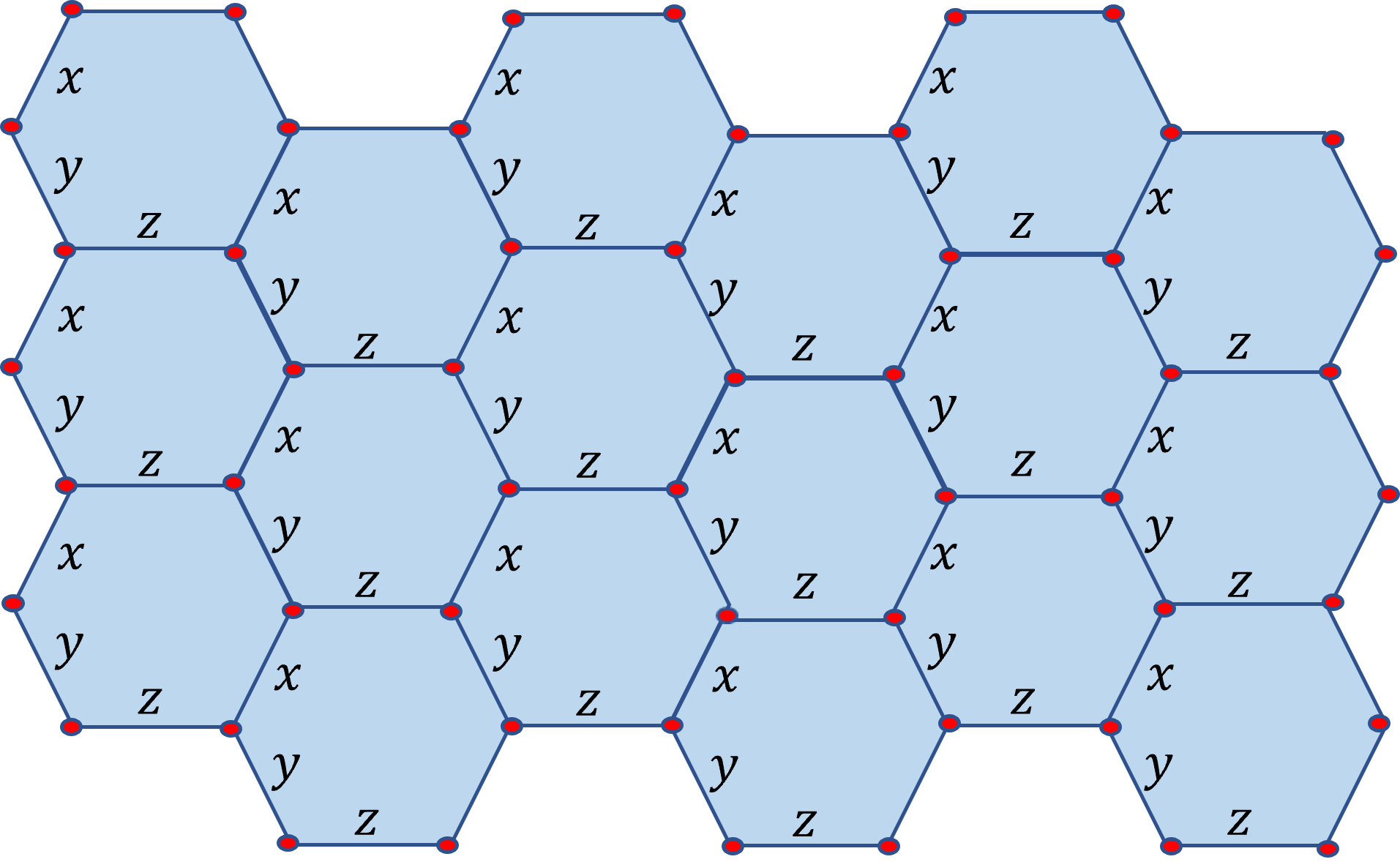}
        \caption{}
        \label{fig:honeycomb lattice}
    \end{subfigure}
    \hfill
    \begin{subfigure}{0.3\textwidth}
        \centering
        \includegraphics[width=\linewidth]{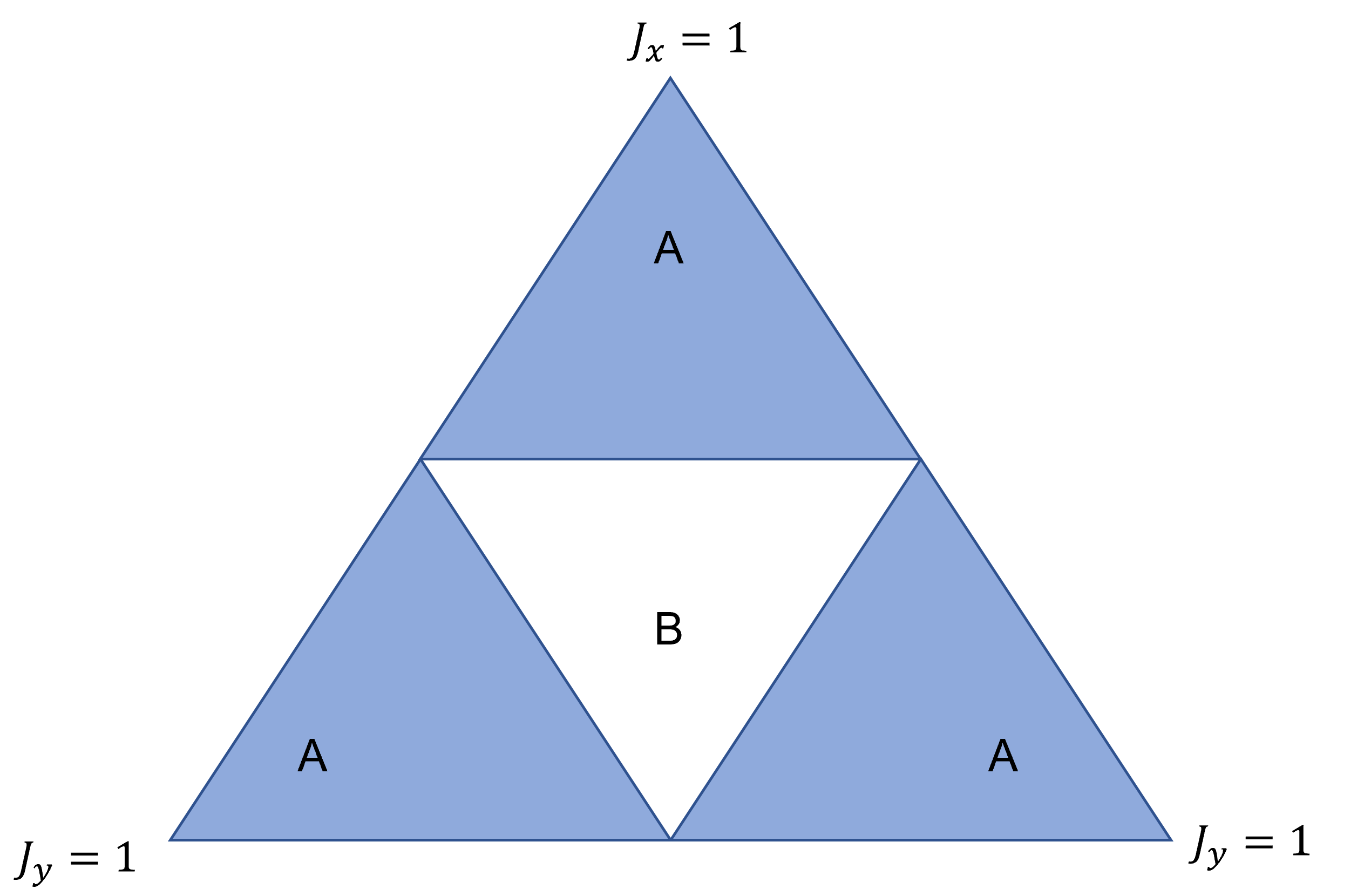}
        \caption{}
        \label{fig: honecomb phase diagram}
    \end{subfigure}
    \hfill
    \begin{subfigure}{0.3\textwidth}
        \centering
        \includegraphics[width=\linewidth]{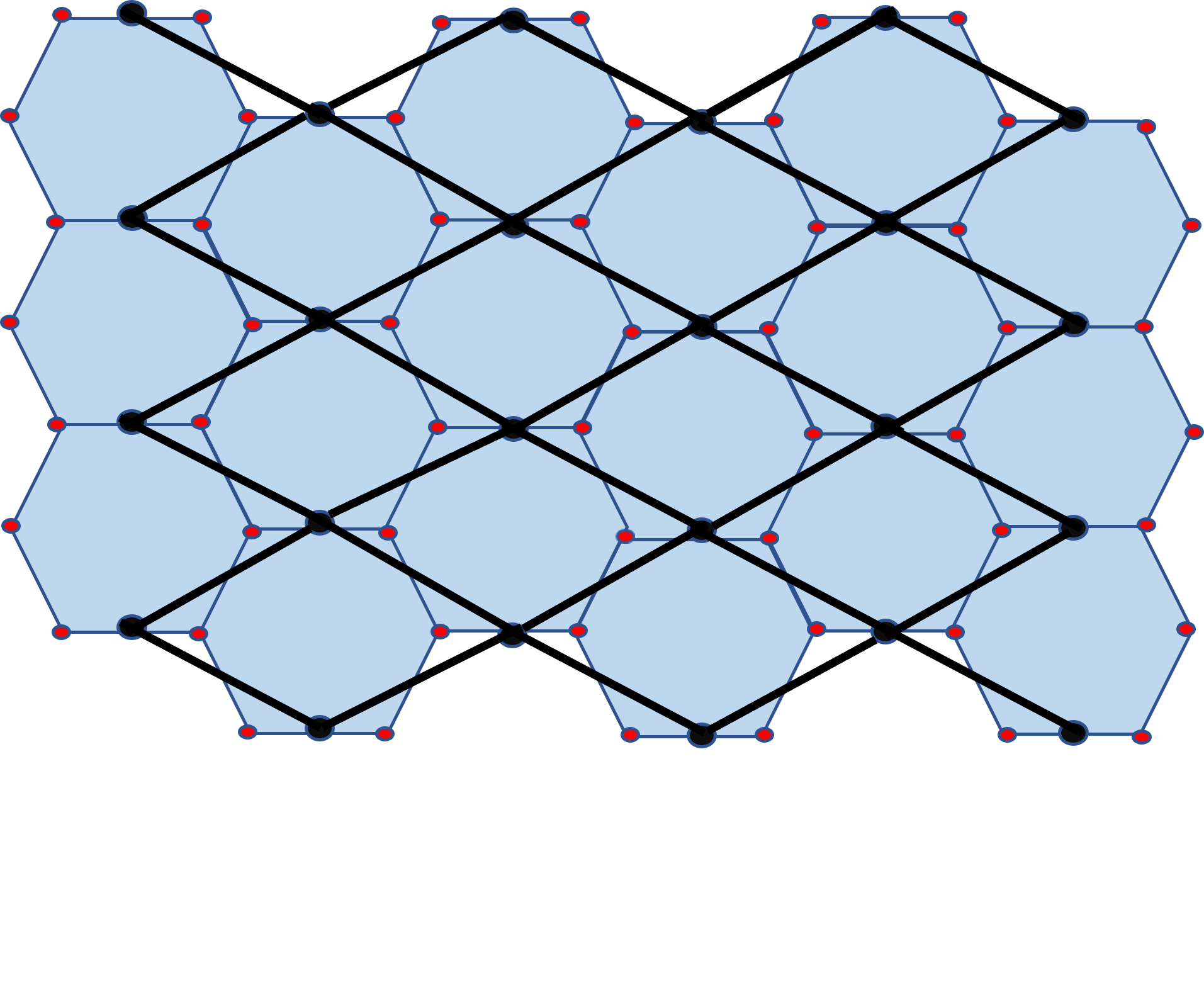}
        \caption{}
        \label{shrinked honeycomb lattice}
    \end{subfigure}
    \caption{(a) Figure 1: The original Honeycomb Lattice.(b) Figure 2: Phase diagram depicting the gapped phases. When one of $J_x$, $J_y$, or $J_z$ is dominant, it is mathematically equivalent to a toric code model. (c) Figure 3: Depiction of the shrunken lattice when $J_z$ is dominant, illustrated by the reduction of two physical qubits to one effective qubit in the ground state of the ``z-edge'' check operators.}
    \label{fig:Honeycomb cases}
\end{figure}

Let us briefly revisit the Kitaev Honeycomb Model and establish some notations. The lattice, depicted in Figure \ref{fig:honeycomb lattice} and denoted by $\Gamma=(V,E,P)$, consists of vertices ($V$), edges ($E$), and plaquettes ($P$).

The notation $\partial_1 e,\ \partial_2 e$ is used to refer to the two vertices at the end of edge $e \in E$, and $\partial e=\{\partial_1 e,\partial_2 e\}$ indicates the set. $N(A)$ is the count of the set $A$. A frequently used symbol $d_v$ denotes the count of set $e|e\in E, v\in \partial e$, i.e., the degree of the vertex $v$. $Bo(p) \subseteq E$, for $p\in P$ represents the edges that border the plaquette $p$.

Each vertex houses a qubit. The total Hilbert Space is:
\begin{equation}
\centering
    \mathcal{H} := \bigotimes_{e \in E} \mathcal{H}_e
\end{equation}
Each edge on the lattice is associated with a symbol $x,y,z$. For the honeycomb lattice, we label all the edges as illustrated in Figure \ref{fig:honeycomb lattice}, consistent with the original paper\cite{Kitaev_2006}. These edges are referred to as ``x-edges'', ``y-edges'', and ``z-edges''. The edge associated with $x(y,z)$ involves a two-body Pauli operator $X\otimes X(Y \otimes Y,Z\otimes Z)$ acting on the qubits at the ends of the edge. The operators linked to edges are defined as check operators, denoted by $P_e$. We say that two check operators are unconnected if the edges associated with these operators are not connected. In this paper, we use $X,Y,Z$ to represent Pauli Operators $\sigma_x,\sigma_y,\sigma_z$:
\[
\sigma_x = \begin{pmatrix} 0 & 1 \\ 1 & 0 \end{pmatrix}, \quad
\sigma_y = \begin{pmatrix} 0 & -i \\ i & 0 \end{pmatrix}, \quad
\sigma_z = \begin{pmatrix} 1 & 0 \\ 0 & -1 \end{pmatrix}
\]

The Hamiltonian is the summation of weighted check operators:
\begin{equation}
    \centering
    H=-J_x \sum_{x-edges} X\otimes X-J_y \sum_{y-edges} Y\otimes Y-J_z \sum_{Z-edges} Z\otimes Z
\end{equation} 

As depicted in Figure \ref{fig: honecomb phase diagram}, the phase diagram of the Honeycomb model is well-defined. In the $A(B)$ region, it represents a gapped(gapless) phase. Kitaev explicitly demonstrated, using perturbation theory, that the gapped phase is the toric code phase, where one of the $J_x,J_y,J_z$ variables is much larger than others. Then those two qubits connected by $z-edges$ will stay at the ground state of the check operator $Z\otimes Z$. We say these two qubits are effectively ``shrunk'' to a single qubit. Subsequently, the lattice is also shrunk by replacing the edge to a single vertex, as shown in Figure \ref{shrinked honeycomb lattice}. This is referred to as the ``shrunken'' lattice.

Let us rephrase this in our notations. We denote these check operators associated with $z-edges$ as a Stabilizer center $S_c$. We obtain a shrunken lattice when this $S_c$ is dominant. The ground state under this limit is twofold: it is simultaneously the ground state of $S_c$ and the ground state of all plaquette terms $W_p$, where $W_p$ is the product of check operators associated to edges in $Bo(p)$.

Generally, on a trivalent lattice, a Hamiltonian of the following form can be considered:
\begin{equation}
    \centering
    H=- \sum_{e \in E} J_e P_e
\end{equation} 
The difinition of check operators could be varied as long as the following commutative relation remains:
\begin{align}
    [P_e,P_{e'}]& =0 \;\; if \;\; \partial e \cap \partial e' = \emptyset\\
     \{P_e,P_{e'}\}&=0 \;\; if \;\; \partial e \cap \partial e' \neq \emptyset\\
\end{align}
For any $e \neq e'$. This means that check operators should anticommute if they intersect at exactly one vertex and commute in other scenarios. We require all check operators in $S_c$ to be unconnected. When we allow $S_c$ to be dominant, we obtain a shrunken lattice by replacing the edges of $S_c$ by a single vertex. We will demonstrate that this will be a specific surface code. Indeed, the shrunken lattice at different $S_c$ may vary. In this honeycomb lattice, the shrunken lattice at dominant $x-edges$ and $z-edges$ are all square lattices. However, in \cite{Hastings_Haah_2021}, their shrunken lattice is a kagome lattice (when the qubits are considered to be placed on vertices). We will frequently utilize the concept of a shrunken lattice at a given $S_c$.

\section{Generalized Method}
\label{sec:generalization}

\subsection{Toric Code on a lattice where qubits are placed on vertices} \label{sec: rewrite the toric code}

\begin{figure}[htbp]
    \centering
    \begin{subfigure}{0.3\textwidth}
        \centering
        \includegraphics[width=\linewidth]{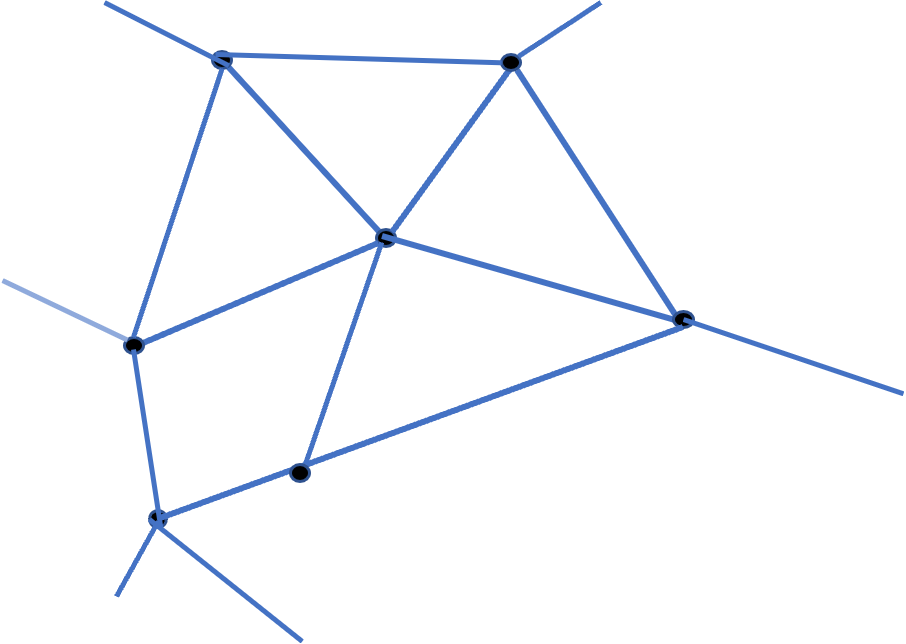}
        \caption{}
        \label{fig:sub1}
    \end{subfigure}
    \hfill
    \begin{subfigure}{0.3\textwidth}
        \centering
        \includegraphics[width=\linewidth]{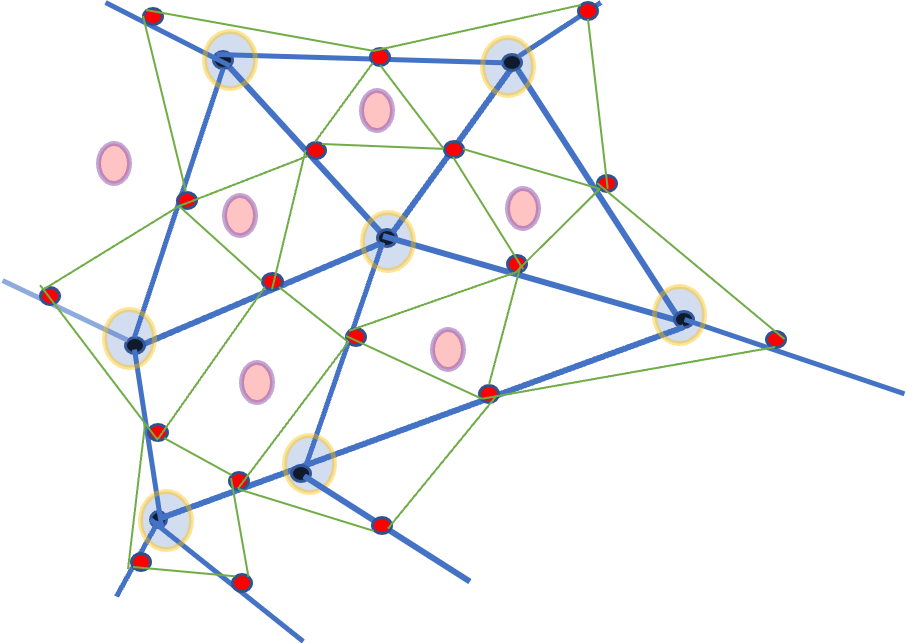}
        \caption{}
        \label{fig:sub2}
    \end{subfigure}
    \hfill
    \begin{subfigure}{0.3\textwidth}
        \centering
        \includegraphics[width=\linewidth]{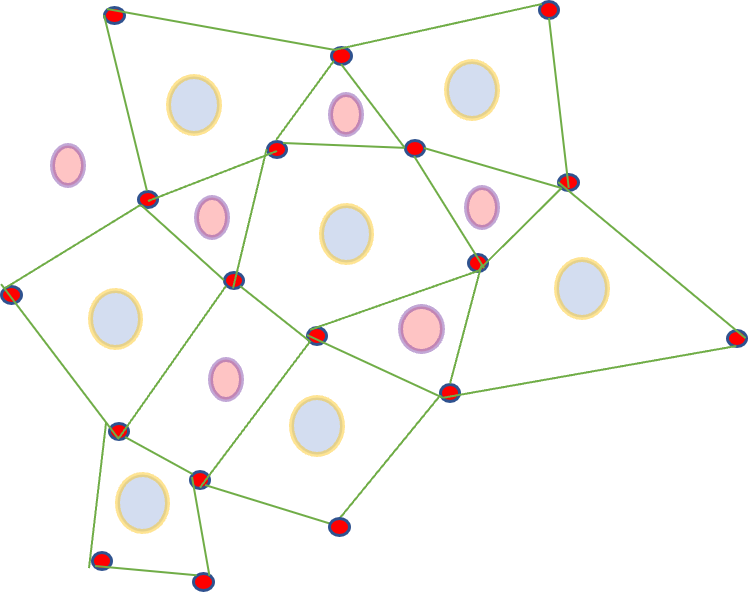}
        \caption{}
        \label{fig:sub3}
    \end{subfigure}
    \caption{(a) Figure 1: Original lattice model with qubits placed on edges. Black dots represent vertices $E$.(b) Figure 2: Transformation process of the lattice. Red dots label the center of edges. Two red dots are connected if they belong to the same plaquette and are  connected. New plaquettes are colored in red and grey circles.(c) Figure 3: The transfromed lattice with qubits on vertices. And the original lattice is removed. $A_v$ and $B_p$ operators act on the two types of plaquettes, labeled by red and grey circles respectively.}
    \label{fig:relattice}
\end{figure}

The toric code model is defined on an arbitrary planar lattice $\Gamma=(V,E,P)$, with one qubit placed on each edge. The Hamiltonian is:

\begin{equation}
    H=-\sum_v A_v-\sum_p B_p
\end{equation}
Here, $A_v=\bigotimes_{{e|v\in \partial e }} X_e$ and similarly, $B_p=\bigotimes_{{e|e\in Bo(p)}} Z_e$. The symbols $X_e$ and $Z_e$ indicate that the Pauli operator $X$ and $Z$ acts on the qubit placed on the edge $e$.
For our purposes, we need to reshape the lattice into a more convenient form as shown in figure \ref{fig:relattice}. The process is as follows. First, we attach a new vertex on each edge $e$, denoted by red dots. We add one edge to connect red dots on $e_1$ and $e_2$ if they satisfy:
\begin{align}
        & e_1 \neq  e_2 \\
       & N(\partial e_1 \cap \partial e_2)   =1 \\
       & \exists p \in P,
 {e_1,e_2} \subseteq Bo(p) 
\end{align}
We will add two edges to connect $e_1$ and $e_2$ if $N(\partial e_1 \cap \partial e_2)   = 2$ in the above requirement. This results in a new lattice $\Gamma'=(V',E',P')$, where $V'$ is the set of red dots and $V'=E$ as sets. $E'$ is the set of newly added edges connecting red dots and $P'=V \cup P$ as sets.

Notably, the degree of the new vertex is automatically $4$. The new plaquettes are two-colored by vertices and plaquettes of $\Gamma$. Consequently, the toric code becomes a lattice model on $\Gamma'$, with one qubit placed on each vertex.  In the new lattice as in figure \ref{fig:sub3}, the plaquette $p_g$($p_r$) with a green(red) circle has a plaquette term that is $\bigotimes X(Z)$ on each qubit on the boundary of $p_g$($p_r$), corresponding to previous $A_v(B_p)$ operators.

\subsection{Generalized model on a lattice with all vertices having even degree}
\label{sec:even codimension}
Begin with a lattice $\Gamma=(V,E,P)$, where each vertex $v \in V$ has an even degree $d_v$. Our purpose is to define check operators and ensure the Hamiltonian remains the weighted summation of these check operators. The check operator $P_e$ should take a form that is a tensor product of Pauli operators acting on the ends of edge $e$, namely $P_e = P_{\partial_1 e} \otimes P_{\partial_2 e}$. To maintain the property that check operators anticommute if they are connected, we require additional operators that anticommute with each other when the vertex has a higher degree. We find that placing $d_v/2 - 1$ qubits on each vertex facilitates this. For $k$ qubits, we have $2k + 1$ mutually anti-commuting Pauli operators as follows:

\begin{equation}
  \begin{array}{ccccccc}
    p_1 &= &1 \otimes & 1\otimes & 1 \otimes & \cdots & X \\
    p_2 &= &1 \otimes & 1 \otimes & 1 \otimes & \cdots & Y \\
    &\vdots & \vdots & \ddots & \vdots \\
    p_{2t+1} &= &\underbrace{1  \cdots \otimes 1 }_{k-t-1} \otimes& X \otimes & Z\otimes & \cdots & Z \\
    p_{2t+2} &= &\underbrace{1  \cdots \otimes 1 }_{k-t-1} \otimes& Y \otimes & Z\otimes & \cdots & Z \\
    &\vdots & \vdots & \ddots & \vdots \\
    P_{2k+1} &= &Z\otimes &Z \otimes &Z \otimes& \cdots & Z
  \end{array}
\end{equation}

Importantly, because we ultimately aim to reach the toric code, the signs of each term do not significantly matter, as different sign configurations are related by unitary transformations. This allows us to consider each operator within the Pauli group $\mathcal{P} = \{G / \{+1, -1, +i, -i\}\}$, where $G$ is designated to represent the set of all possible tensor products of Pauli operators.

Within the Pauli group, the phase gate $P_{\text{gate}}$ interchanges $X$ and $Y$, while leaving $Z$ unaffected. This can swap $P_{2t+1}$ with $P_{2t+2}$ for any $0 \leq t < k$. Subsequently, the Hadamard gate $H_{\text{gate}}$ flips $X$ and $Z$, which in turn flips $P_{2k+1}$ with $P_{2k-1}$. The gate $S_{\text{gate}} = P_{\text{gate}} \circ H_{\text{gate}} \circ P_{\text{gate}}$ flips $Y$ and $Z$. CNOT gate is a unitary operator. Elementary actions by conjugating CNOT gate on two-qubits Pauli operators are given by: 
\begin{equation}
\begin{aligned}
\text{CNOT}(IX) &= IX, & \text{CNOT}(XI) &= XX, \\
\text{CNOT}(IZ) &= ZZ, & \text{CNOT}(ZI) &= ZI.
\end{aligned}
\end{equation}

A sequence of conjugations of  operators above is then sufficient to flip $P_1$ with $P_4$, noting that it is sufficient to consider only the last two qubits. Here is how to flip $X \otimes Z$ with $1 \otimes Y$, without changing the other operators:

\[
\begin{aligned}
X \otimes Z &\xrightarrow{H_{gate} \otimes \text{Id}} &Z \otimes Z &\xrightarrow{\text{CNOT}} &1 \otimes Z &\xrightarrow{S_{gate} \otimes S_{gate}} &1 \otimes Y &\xrightarrow{P_{gate} \otimes 1} &1 \otimes Y\\
 1 \otimes Y &\xrightarrow {H_{gate} \otimes \text{Id}}& 1 \otimes Y &\xrightarrow{\text{CNOT}} &Z \otimes Y &\xrightarrow{S_{gate} \otimes S_{gate}}& Y \otimes Z &\xrightarrow{P_{gate} \otimes 1}& X \otimes Z
\end{aligned}
\]

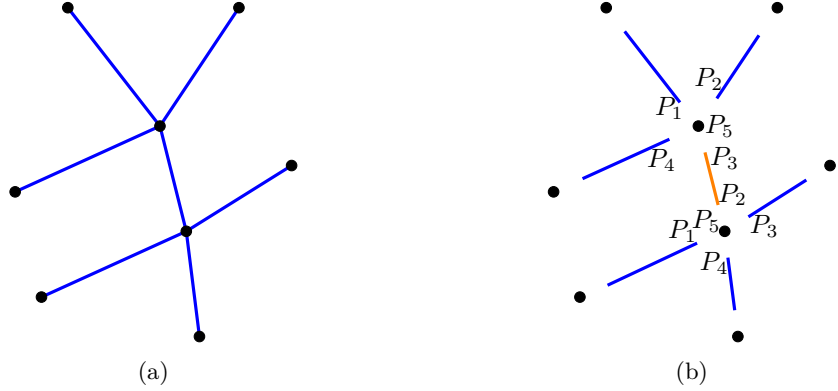
\begin{figure}[htbp]
    \centering
    \begin{subfigure}{0.4\textwidth}
        \centering
        \begin{tikzpicture}[scale=0.35]
        \draw[very thick, blue] (0,0) -- (-3.5,4.5);
        \draw[very thick, blue] (0,0) -- (3,4.5);
        \draw[very thick, blue] (0,0) -- (-5.5,-2.5);
        \draw[very thick, blue] (0,0)-- (1,-4);
        \draw[very thick, blue] (1,-4)-- (5,-1.5);
        \draw[very thick, blue] (1,-4)-- (1.5,-8);
        \draw[very thick, blue] (1,-4)-- (-4.5,-6.5);
        
        \filldraw[black] (-3.5,4.5) circle (0.2);
        \filldraw[black] (3,4.5) circle (0.2);
        \filldraw[black] (-5.5,-2.5) circle (0.2);
        \filldraw[black] (0,0) circle (0.2);
        \filldraw[black] (1,-4) circle (0.2);
        \filldraw[black] (5,-1.5) circle (0.2);
        \filldraw[black] (1.5,-8) circle (0.2);
        \filldraw[black] (-4.5,-6.5) circle (0.2);
        \end{tikzpicture}
        \caption{}
        \label{fig:d_v=4 local 1}
    \end{subfigure}
    \hfill
    \begin{subfigure}{0.4\textwidth}
        \centering
        \begin{tikzpicture}[scale=0.35]
        \draw[very thick, blue] (-0.7,0.9) -- (-2.8,3.6);
        \draw[very thick, blue] (0.7,1.05) -- (2.3,3.45);
        \draw[very thick, blue] (-1.1,-0.5) -- (-4.4,-2);
        \draw[very thick, orange] (0.25,-1) -- (0.75,-3);
        \draw[very thick, blue] (1.9,-3.45) -- (4.1,-2.05);
        \draw[very thick, blue] (1.12,-5) -- (1.38,-7);
        \draw[very thick, blue] (-0.05,-4.45) -- (-3.45,-6.05);
        \filldraw[black] (-3.5,4.5) circle (0.2);
        \filldraw[black] (3,4.5) circle (0.2);
        \filldraw[black] (-5.5,-2.5) circle (0.2);
        \filldraw[black] (0,0) circle (0.2);
        \filldraw[black] (1,-4) circle (0.2);
        \filldraw[black] (5,-1.5) circle (0.2);
        \filldraw[black] (1.5,-8) circle (0.2);
        \filldraw[black] (-4.5,-6.5) circle (0.2);
        \node[black] at (0.8,0) {$P_{5}$};
        \node[black] at (0.4,1.8) {$P_{2}$};
        \node[black] at (-1.1,0.7) {$P_{1}$};
        \node[black] at (-1.4,-1.3) {$P_{4}$};
        \node[black] at (1,-1.2) {$P_{3}$};
        \node[black] at (0.3,-3.6) {$P_{5}$};
        \node[black] at (1.3,-2.6) {$P_{2}$};
        \node[black] at (-0.6,-4) {$P_{1}$};
        \node[black] at (0.6,-5.2) {$P_{4}$};
        \node[black] at (2.5,-3.8) {$P_{3}$};
        \end{tikzpicture}
        \caption{}
        \label{fig:d_v=4 local 2}
    \end{subfigure}
    \hfill

    \caption{(a) Figure 1: A local part of the entire lattice diagram. (b) Figure 2: A simplified illustration of the assignment of Pauli operator $P_1$ through $P_4$ from each vertex $v$ to the surrounding $P_e$, while $P_5$ is assigned to the $P_v$. Each $P_e$ is the tensor product of operators from the two end vertices of $e$. For example, the operator on the orange edge is $(1 \otimes X) \otimes (Y \otimes Z)$, or simply $P_3 \otimes P_2$.}
    \label{fig:unit of generalization}
\end{figure}

This approach is sufficient to exchange any $P_i$ with $P_j$ by stacking the aforementioned operations, asserting that any distribution of these operators is equivalent. Consider a lattice where all vertices have a degree of four. The check operator on an edge, $P_e$, can be defined as the tensor product of Pauli operators supported on the vertices at the end of the edge $e$. It is important to note that the actual assignment of a Pauli operator for one $P_e$ is not crucial, as long as $P_e$ operators anticommute with each other when they are connected. Figure~\ref{fig:unit of generalization} provides an example of the assignment of Pauli operators, and any other assignment is equivalent up to a unitary transformation. An operator $P_v$ represents a new type of check operator that is associated with only one vertex $v$, which, for our convenience, is chosen as $P_v = P_5$. The Hamiltonian is as follows:

\begin{align}
\centering
    H=-\sum_{v} J_v P_v-\sum_{e} J_e P_e
\end{align}
Now let $J_v$ dominate. Note that these operators commute with each other, hence they share common eigenspaces. Let $J_e \ll J_v$, with $J_v = 1$, and examine the corresponding perturbation theory where $H_0 = \sum_v P_v$ and $H' = \lambda \sum_{e} J_e P_e$, the perturbation Hamiltonian. Here, $\lambda$ is a small factor to denote the perturbation order. Denote $|GS\rangle$ as the ground state of $H_0$. As in Kitaev's paper \cite{Kitaev_2006}, the effective Hamiltonian around the ground state is given by:
\begin{equation}
    H_{\text{eff}} = T \left(H' + H' G_0' H' + H' G_0' H' G_0' H' + \ldots\right)T
\end{equation}
where $T = |GS\rangle \langle GS|$ is the projector onto the ground state of $H_0$, and $G_0' = \left(\frac{1}{E_0 - H_0}\right)'$ is the Green's function, where the prime notation implies that $G_0'$ vanishes on the ground state and acts normally on the excited states.

Appendix~\ref{appendix:perturbation theory} provides an explicit treatment of the perturbation method; here, we derive the effective Hamiltonian:

\begin{equation} \label{eqn: effective Hamiltonian formula}
    H_{eff}= (-1)^{\gamma_p}\sum_p \alpha_{p} \lambda^{l_p} W_p +  constant
\end{equation}
$W_p$ is the plaquette operator, which is the product of check operators bordering the plaquette. $l_p$ indicates the perturbation order and $(-1)^{\gamma_p}$ is used to fullfill the gap between the perturbed effective Hamiltonian with $W_p$. They are explained in the appendix. $\alpha_{p}$ is an interesting path-dependent factor arises from the perturbation and we leave an interesting discussion of the zero point property in the appendix~\ref{appendix:perturbation theory}.

\begin{figure}[htbp] 
    \centering
    \begin{subfigure}{0.35\textwidth}
        \centering
        \begin{tikzpicture}[scale=0.35]
        \draw[thick, black] (-4,-4) -- (4,4);
        \draw[thick, black] (-4,4) -- (4,-4);
        \node[black] at (1.5,0) {\footnotesize $1Z$};
        \node[black] at (-1.5,0) {\footnotesize $Z1$};
        \node[black] at (0,1.5) {\footnotesize $YY$};
        \node[black] at (0,-1.5) {\footnotesize $XX$};
        \node[red] at (5,0) {\footnotesize $Z$};
        \node[red] at (-5,0) {\footnotesize $Z$};
        \node[red] at (0,5) {\footnotesize $X$};
        \node[red] at (0,-5) {\footnotesize $X$};
        \node[LimeGreen] at (4.2,4.4) {\footnotesize $ZY$};
        \node[LimeGreen] at (-4.2,4.4) {\footnotesize $X1$};
        \node[LimeGreen] at (4.2,-4.4) {\footnotesize $ZX$};
        \node[LimeGreen] at (-4.2,-4.4) {\footnotesize $Y1$};
        \end{tikzpicture}
        \caption{}
        \label{fig:d_v=4 case}
    \end{subfigure}
    \hfill
    \begin{subfigure}{0.6\textwidth}
        \centering
        \begin{tikzpicture}[scale=0.35]
        \draw[thick, black] (13,3.2) -- (15,3.2);
        \draw[thick, black] (13,3.8) -- (15,3.8);
        \draw[thick, black] (14.5,4.15) -- (15.5,3.5) -- (14.5,2.85);
        
        \draw[thick, black] (7,3.5) -- (4,6);
        \draw[thick, black] (7,3.5) -- (4,1);
        \draw[thick, black] (7,3.5) -- (7,-0.5);
        \draw[thick, black] (7,3.5) -- (10,1);
        \draw[thick, black] (7,3.5) -- (11,4.5);
        \draw[thick, black] (7,3.5) -- (9,7);
        \node[LimeGreen] at (9.2,7.8) {\footnotesize $ZY$};
        \node[LimeGreen] at (9.2,7.2) {\footnotesize $1$};
        \node[LimeGreen] at (11.5,4.8) {\footnotesize $ZZ$};
        \node[LimeGreen] at (11.5,4.2) {\footnotesize $X$};
        \node[LimeGreen] at (10.5,0.8) {\footnotesize $ZZ$};
        \node[LimeGreen] at (10.5,0.2) {\footnotesize $Y$};
        \node[LimeGreen] at (7,-1) {\footnotesize $ZX$};
        \node[LimeGreen] at (7,-1.6) {\footnotesize $1$};
        \node[LimeGreen] at (3.5,0.8) {\footnotesize $Y1$};
        \node[LimeGreen] at (3.5,0.2) {\footnotesize $1$};
        \node[LimeGreen] at (3.5,6.8) {\footnotesize $X1$};
        \node[LimeGreen] at (3.5,6.2) {\footnotesize $1$};
        \node[red] at (6.3,7.5) {\footnotesize $XZ$};
        \node[red] at (11,6.5) {\footnotesize $1X$};
        \node[red] at (11,2.5) {\footnotesize $1Z$};
        \node[red] at (8.8,-0.3) {\footnotesize $ZX$};
        \node[red] at (5.2,-0.3) {\footnotesize $X1$};
        \node[red] at (3.2,3.5) {\footnotesize $Z1$};
        \node[black] at (5.6,3.75) {\scriptsize $Z1$};
        \node[black] at (5.6,3.25) {\scriptsize $1$};
        \node[black] at (6.25,2.1) {\scriptsize $XX$};
        \node[black] at (6.25,1.6) {\scriptsize $1$};
        \node[black] at (7.7,2.1) {\scriptsize $1Y$};
        \node[black] at (7.7,1.6) {\scriptsize $Y$};
        \node[black] at (6.8,4.8) {\scriptsize $YY$};
        \node[black] at (6.8,4.3) {\scriptsize $1$};
        \node[black] at (8.8,4.6) {\scriptsize $1XX$};
        \node[black] at (8.7,3.2) {\scriptsize $11Z$};
        
        \draw[thick, black] (17,3.5) -- (23,3.5);
        \draw[thick, black] (18.7,1.5) -- (18.7,5.5);
        \draw[thick, black] (21.3,1.5) -- (21.3,5.5);
        \node[red] at (18.2,4) {\footnotesize $X$};
        \node[red] at (18.2,3) {\footnotesize $Z$};
        \node[red] at (19.2,4) {\footnotesize $Z$};
        \node[red] at (19.2,3) {\footnotesize $X$};
        \node[red] at (21.8,4) {\footnotesize $X$};
        \node[red] at (21.8,3) {\footnotesize $Z$};
        \node[red] at (20.8,4) {\footnotesize $Z$};
        \node[red] at (20.8,3) {\footnotesize $X$};
        \end{tikzpicture}
        \caption{}
        \label{fig:d_v=6 case}
    \end{subfigure}
    \hfill

    \caption{(a) Figure 1: This illustration demonstrates how the effective plaquette terms are obtained on the ground state of the $P_e$s. Each operator in the figure represents the action of either a check operator or a plaquette operator on the qubits located at the vertices. The green operator represents one of the anticommuting Clifford operators associated with edges. The black operator illustrates the plaquette term on a given vertex, and the red operator presents the effective plaquette term on the same vertex. (b) Figure 2: This depiction also shows how a vertex with degree $d_v=6$ is transformed into two connected vertices, each with a degree of $d_v=4$. It is essential to note that the effective action is consistent with the toric code case.}
    \label{fig:even vertex}
\end{figure}
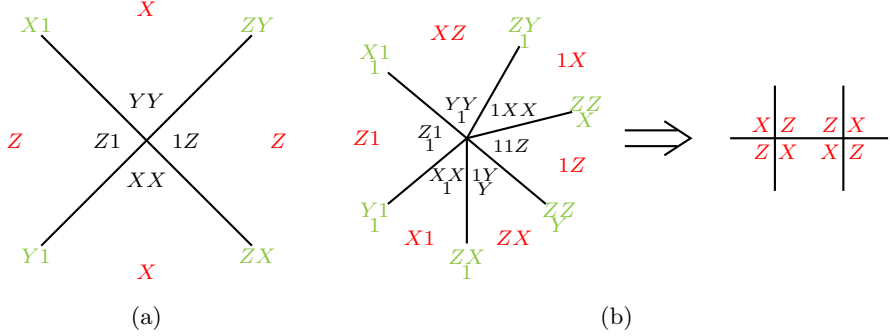

When $P_v$ is dominant, the two qubits placed on the vertex effectively become one qubit. The corresponding shrunken lattice, illustrated in Figure~\ref{fig:d_v=4 case}, has $d_v=4$, and the action of plaquette terms $W_p$ around each vertex exerts the same local action on the vertex as in the toric code case, as shown in Figure~\ref{sec: rewrite the toric code}.

Vertices with $d_v=2k$ can be treated similarly, where $k>2$ and $k$ is an integer. Generally, for a vertex with an even $d_v$, we place $k=d_v/2 - 1$ qubits on the vertex, and designate $\bigotimes_{k} Z$, or $P_{2k+1}$, to be $P_v$ and distribute the remaining $d_v$ Pauli operators to the surrounding edges. The check operator is defined similarly to the case where $d_v=4$. Then, the phase we are investigating is when all $P_v$s are dominant. We illustrate the example of a vertex with $d_v=6$ in Diagram~\ref{fig:d_v=6 case}. The computation is grounded on the mapping table to find the effective Hamiltonian, as in Table~\ref{tab:mapping table for a vertex with $d_v=6$ in 2d}. Essentially, we provide a specific distribution of operators around the vertex and calculate the effective action of the plaquette terms on this vertex. We observe that the effective two qubits split into two connected vertices with $d_v=4$. It is clear that both vertices maintain consistent and identical local properties as of the toric code. The generic mapping table for a vertex with degree $d_v \geq 4$ is shown in Appendix~\ref{generic mapping table}. When we examine the ground state of dominant $P_v$s, each vertex with degree $d_v=2k$ will split into $k-1$ vertices with degree 4.

\begin{table}[h] % 'h'表示将表格放在此处（适当的位置）
\centering % 居中表格

\begin{tabular}{ll} % 3列，分别为左对齐、居中对齐和右对齐
\toprule
Operator & Effective operator  \\
\midrule
$X \otimes X \otimes 1 $ & $X \otimes 1 $ \\
$1 \otimes X \otimes X$ & $1 \otimes X$  \\
$1 \otimes 1 \otimes Z$ & $1 \otimes Z$  \\
$1 \otimes Z \otimes 1$ & $Z \otimes Z$  \\
$Z \otimes 1 \otimes 1$ & $Z \otimes 1$ \\
\bottomrule
\end{tabular}
\caption{Mapping Table for a vertex with $d_v=6$} % 表格标题
\label{tab:mapping table for a vertex with $d_v=6$ in 2d} % 用于引用的标签

\end{table}
After splitting all vertices with $d_v=2k$, we obtain the shrunken lattice. We conclude that a $\mathbf{Z_2}$ phase is recovered in the generalized Kitaev model with even degree vertices when the shrunken lattice is two-colorable.

\subsection{Generalized model on arbitrary planar lattice}\label{sec:on general lattice}
The remaining question concerns how to address vertices with an odd degree $d_v \geq 5$. It is logical to place $(d_v-1)/2$ qubits on each vertex and distribute $d_v$ Clifford operators to the surrounding edges so that all check operators are defined. The general Hamiltonian on a lattice $\Gamma = (V, E, P)$ is given by:
\begin{align}\label{eqn:original Hamiltonian}
    H &= -\sum_{\{v | d_v = 2k, k \in \mathbb{Z}\}} J_v P_v - \sum_{e \in E} J_e P_e,\\
    P_e &= P_{\partial_1 e} \otimes P_{\partial_2 e}.
\end{align}

We designate the stabilizer center $S_c$ as $H_0$, which includes all $\{P_v | v \in V\}$ and a subset $S_e$ of $\{P_e | e \in E\}$ such that any two $P_e \in S_e$ commute with each other and with $\{P_v | v \in V\}$. In other words, $S_e$ consists of check operators on the edges that connect vertices with an odd $d_v$. If we further require that the shrunken lattice at $S_c$ is a two-colorable degree-4 lattice, or equivalently, that any two vertices with an odd degree are shrunk, then the effective Hamiltonian resembles the toric code model when the coefficients of $S_c$ are dominant.

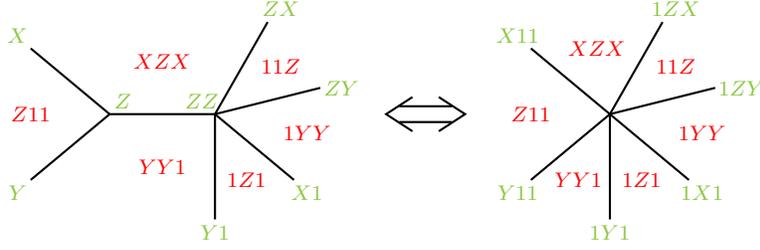
\begin{figure}[htbp]
\centering
\begin{tikzpicture}[scale=0.35]
\draw[thick, black] (-1,3.2) -- (1,3.2);
\draw[thick, black] (-1,3.8) -- (1,3.8);
\draw[thick, black] (0.5,4.15) -- (1.5,3.5) -- (0.5,2.85);
\draw[thick, black] (-0.5,4.15) -- (-1.5,3.5) -- (-0.5,2.85);

\draw[thick, black] (7,3.5) -- (4,6);
\draw[thick, black] (7,3.5) -- (4,1);
\draw[thick, black] (7,3.5) -- (7,-0.5);
\draw[thick, black] (7,3.5) -- (10,1);
\draw[thick, black] (7,3.5) -- (11,4.5);
\draw[thick, black] (7,3.5) -- (9,7);
\node[LimeGreen] at (9.5,7.5) {\footnotesize $1ZX$};
\node[LimeGreen] at (12,4.5) {\footnotesize $1ZY$};
\node[LimeGreen] at (10.5,0.5) {\footnotesize $1X1$};
\node[LimeGreen] at (7,-1) {\footnotesize $1Y1$};
\node[LimeGreen] at (3.5,0.5) {\footnotesize $Y11$};
\node[LimeGreen] at (3.5,6.5) {\footnotesize $X11$};
\node[red] at (6.5,6) {\footnotesize $XZX$};
\node[red] at (9.5,5.3) {\footnotesize $11Z$};
\node[red] at (10.5,2.8) {\footnotesize $1YY$};
\node[red] at (8.2,1) {\footnotesize $1Z1$};
\node[red] at (5.8,1) {\footnotesize $YY1$};
\node[red] at (4,3.5) {\footnotesize $Z11$};

\draw[thick, black] (-8,3.5) -- (-8,-0.5);
\draw[thick, black] (-8,3.5) -- (-5,1);
\draw[thick, black] (-8,3.5) -- (-4,4.5);
\draw[thick, black] (-8,3.5) -- (-6,7);
\node[LimeGreen] at (-5.5,7.5) {\footnotesize $ZX$};
\node[LimeGreen] at (-3.2,4.5) {\footnotesize $ZY$};
\node[LimeGreen] at (-4.5,0.5) {\footnotesize $X1$};
\node[LimeGreen] at (-8,-1) {\footnotesize $Y1$};
\node[red] at (-5.5,5.3) {\footnotesize $11Z$};
\node[red] at (-4.5,2.8) {\footnotesize $1YY$};
\node[red] at (-6.8,1) {\footnotesize $1Z1$};

\draw[thick, black] (-8,3.5) -- (-12,3.5);
\draw[thick, black] (-15,6) -- (-12,3.5);
\draw[thick, black] (-15,1) -- (-12,3.5);
\node[LimeGreen] at (-15.5,0.5) {\footnotesize $Y$};
\node[LimeGreen] at (-15.5,6.5) {\footnotesize $X$};
\node[LimeGreen] at (-11.5,4) {\footnotesize $Z$};
\node[LimeGreen] at (-8.5,4) {\footnotesize $ZZ$};
\node[red] at (-15,3.5) {\footnotesize $Z11$};
\node[red] at (-10,5.5) {\footnotesize $XZX$};
\node[red] at (-10,1.5) {\footnotesize $YY1$};
\end{tikzpicture}
\caption{This illustration demonstrates that a vertex of degree 3 combined with a vertex of degree 5 is equivalent to a single vertex of degree 6. In the left above, a $P_e=Z \otimes Z \otimes Z$ over the red edge is put into the $S_c$, and in the right above, a $P_v=Z \otimes Z \otimes Z$ is put into the $S_c$. They have the  same action of surrounding plaquette, hence these two cases are equivalent for our purpose.}
\label{fig:2 odd to one even vertex}
\end{figure}

The proof involves transforming the lattice into one where all vertices have even degrees. Notably, if we make the operators on the edge connecting two vertices with odd degrees $d_1$ and $d_2$ dominant, it is algebraically equivalent to a single vertex with degree $d_1 + d_2 - 2$ and a dominant $P_v$. An example is illustrated in Figure~\ref{fig:2 odd to one even vertex}, and the general case follows similarly. However, a vertex of odd degree cannot be made equivalent to the combination of two vertices with lower degree, thus making the generalization nontrivial.

We now conclude that on a general lattice $\Gamma$, if there exists a set $S_c$ such that all vertices are shrunk and the resulting shrunk lattice is 2-colorable, the generalized Kitaev Spin liquid model resides in the $\mathbf{Z_2}$ phase.

\section{Emergent Twist Defect in the gapped phase} \label{sec:twist defect}
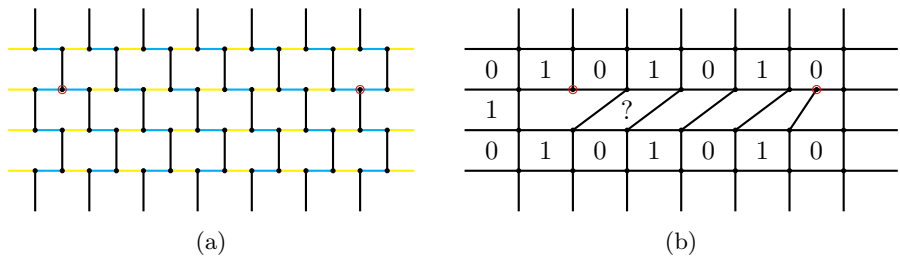
\begin{figure}[htbp]
    \centering
    \begin{subfigure}{0.45\textwidth}
        \centering
        \begin{tikzpicture}[scale=0.18]
        \foreach \i in {2,6,10,14,18,22,26}
        \foreach \j in {3,9,15}
                \draw[thick, black] ({\i},{\j}) -- ({\i},{\j-3});
        \foreach \i in {4,8,12,16,20,24,28}
        \foreach \j in {6,12}
                \draw[thick, black] ({\i},{\j}) -- ({\i},{\j-3});
        \foreach \i in {2,6,10,14,18,22,26,30}
        \foreach \j in {3,6,12}
                \draw[thick, yellow] ({\i},{\j}) -- ({\i-2},{\j});
        \foreach \i in {2,8,12,16,20,24,30}
                \draw[thick, yellow] ({\i},9) -- ({\i-2},9);
        \foreach \i in {2,6,10,14,18,22,26}
        \foreach \j in {3,6,12}
                \draw[thick, cyan] ({\i},{\j}) -- ({\i+2},{\j});
        \foreach \i in {2,4,8,12,16,20,24,26}
                \draw[thick, cyan] ({\i},9) -- ({\i+2},9);
        
        \foreach \i in {2,6,10,14,18,22,26}
        \foreach \j in {3,6,9,12}
                \filldraw[black] ({\i},{\j}) circle (0.15);
        \foreach \i in {4,8,12,16,20,24,28}
        \foreach \j in {3,6,9,12}
                \filldraw[black] ({\i},{\j}) circle (0.15);
        \draw[red] (4,9) circle (0.3);
        \draw[red] (26,9) circle (0.3);
        \end{tikzpicture}
        \caption{}
        \label{fig:lattice with a defect}
    \end{subfigure}
    \hfill
    \begin{subfigure}{0.5\textwidth}
        \centering
        \begin{tikzpicture}[scale=0.18]
        \foreach \i in {4,8,12,16,20,24,28}
                \draw[thick, black] ({\i},0) -- ({\i},6);
        \foreach \i in {4,8,12,16,20,24,28}
                \draw[thick, black] ({\i},9) -- ({\i},15);
        \foreach \j in {3,6,9,12}
                \draw[thick, black] (0,{\j}) -- (32,{\j});
        \foreach \i in {4,8,12,16,20,24,28}
        \foreach \j in {3,6,9,12}
                \filldraw[black] ({\i},{\j}) circle (0.15);
        \foreach \i in {8,12,16,20}
                \draw[thick, black] ({\i},6) -- ({\i+4},9);
        \foreach \i in {4,28}
                \draw[thick, black] ({\i},6) -- ({\i},9);
        \draw[thick, black] (24,6) -- (26,9);
        \filldraw[black] (26,9) circle (0.15);
        \draw[red] (8,9) circle (0.3);
        \draw[red] (26,9) circle (0.3);
        
        \node[black] at (12,7.5) {$?$};
        \foreach \i in {2,10,18,26}
        \foreach \j in {4.5,10.5}
                \node[black] at ({\i},{\j}) {$0$};
        \foreach \i in {6,14,22}
        \foreach \j in {4.5,10.5}
                \node[black] at ({\i},{\j}) {$1$};
        \node[black] at (2,7.5) {$1$};
        \end{tikzpicture}
        \caption{}
        \label{fig:shrinked toric code with defect}
    \end{subfigure}
    \hfill
    
  \caption{(a) Displays a situation where two trivalent vertices (labelled by red circles) are not shrunk with another vertex. (b) shows the underlined phase where all check operators on yellow edges are dominant. This represents the toric code with a pair of defects. }
    \label{fig:lattice with defect}
\end{figure}

In all previous instances, we selected $S_c$ such that vertices with odd degrees were paired with each other, ensuring that any check operator would violate two terms in $S_c$. However, what happens if there is an odd-degree vertex, such as a trivalent vertex, that has not been paired with another odd-degree vertex? Revisiting the Honeycomb lattice, as depicted in Figure~\ref{fig:shrinked toric code with defect}, the effective Hamiltonian resembles a toric code model with two defects, similar to the findings in \cite{Bombin_2010}. Notably, in our case, there is one additional plaquette as well as one more qubit within the defect line. This demonstrates that we can create a dislocation defect at the toric code level with a regular lattice at the spin liquid level! As studied in \cite{PhysRevB.90.134404}, this type of lattice dislocation defect could capture unpaired Majorana modes in the original Honeycomb model.

%Let us calculate $\alpha_{p}$ for these different plaquettes explicitly. For a standard square, $\alpha_{4}=-1/16$, while for a pentagonal plaquette, $\alpha_{5}=-1/16$. The calculation of the parallelogram between the two defects mirrors the situation as a regular square in the Kitaev Honeycomb case, so it also yields a non-zero result. The defect emerges as an Ising anyon exhibiting non-Abelian statistics, as experimentally observed by \cite{andersen_non-abelian_2023}.

It's important to note that the lattice itself is regular and the lattice dislocation at the toric code level is due to a specific choice of $S_c$! Remarkably, altering $S_c$ is simpler than deforming the lattice itself. We will demonstrate in the following that a linearly interpolating Hamiltonian with different $S_c$ choices can manipulate the defects.

\begin{figure}[htbp]
    \centering
    \begin{subfigure}{0.45\textwidth}
        \centering
        \begin{tikzpicture}[scale=0.18]
        \foreach \i in {2,6,10,14,18,22,26}
        \foreach \j in {3,9,15}
                \draw[thick, black] ({\i},{\j}) -- ({\i},{\j-3});
        \foreach \i in {4,8,12,16,20,24,28}
        \foreach \j in {6,12}
                \draw[thick, black] ({\i},{\j}) -- ({\i},{\j-3});
        \foreach \i in {2,6,10,14,18,22,26,30}
        \foreach \j in {3,6,9,12}
                \draw[thick, yellow] ({\i},{\j}) -- ({\i-2},{\j});
        \foreach \i in {2,6,10,14,18,22,26}
        \foreach \j in {3,6,9,12}
                \draw[thick, cyan] ({\i},{\j}) -- ({\i+2},{\j});
        
        \foreach \i in {2,6,10,14,18,22,26}
        \foreach \j in {3,6,9,12}
                \filldraw[black] ({\i},{\j}) circle (0.15);
        \foreach \i in {4,8,12,16,20,24,28}
        \foreach \j in {3,6,9,12}
                \filldraw[black] ({\i},{\j}) circle (0.15);
        \end{tikzpicture}
        \caption{}
        \label{fig:spin liquid generating TC}
    \end{subfigure}
    \hfill
    \begin{subfigure}{0.5\textwidth}
        \centering
        \begin{tikzpicture}[scale=0.18]
        \foreach \i in {4,8,12,16,20,24,28}
                \draw[thick, black] ({\i},0) -- ({\i},15);
        \foreach \j in {3,6,9,12}
                \draw[thick, black] (0,{\j}) -- (32,{\j});
        \foreach \i in {4,8,12,16,20,24,28}
        \foreach \j in {3,6,9,12}
                \filldraw[black] ({\i},{\j}) circle (0.15);
        \end{tikzpicture}
        \caption{}
        \label{fig:shrinked toric code without defect}
    \end{subfigure}
    \hfill
    
  \caption{(a) Shows a honeycomb lattice where all vertices have been paired and shrunk by dominating the yellow edges.
(b) The resulting effective or shrunk lattice where a toric code Hamiltonian acts.}
    \label{fig:lattice without defect}
\end{figure}
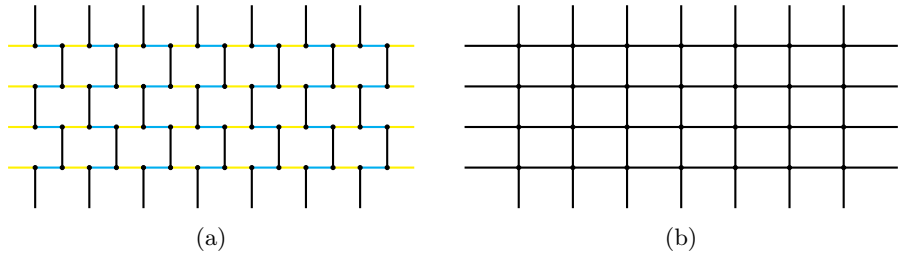

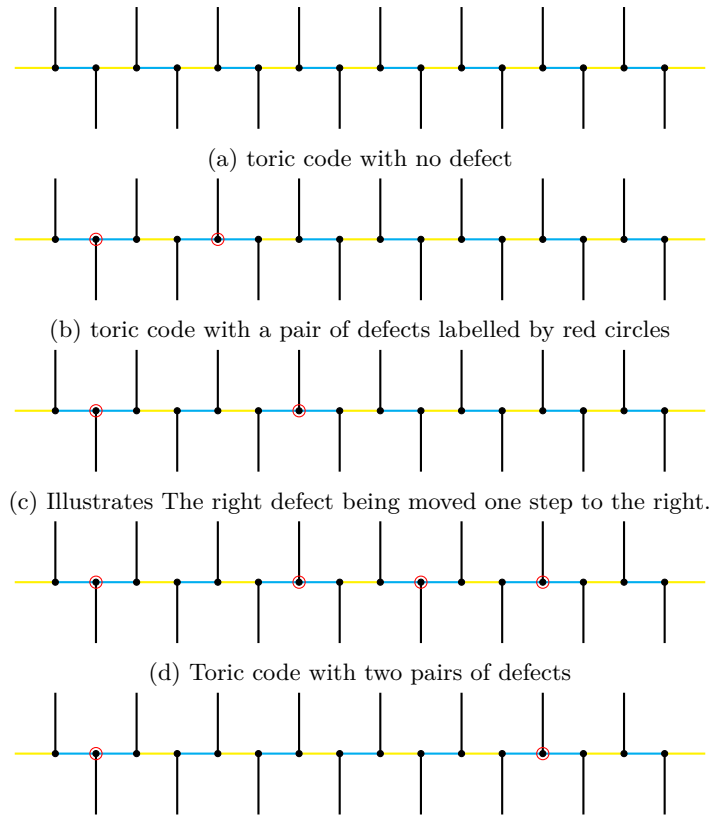
\begin{figure}[htp]
    \centering
    \begin{subfigure}{\textwidth}
        \centering
        \begin{tikzpicture}[scale=0.27]
        \foreach \i in {2,6,10,14,18,22,26,30}
                \draw[thick, black] ({\i},0) -- ({\i},3);
        \foreach \i in {2,6,10,14,18,22,26,30}
                \draw[thick, black] ({\i+2},0) -- ({\i+2},-3);
        \foreach \i in {2,6,10,14,18,22,26,30,34}
                \draw[thick, yellow] ({\i},0) -- ({\i-2},0);
        \foreach \i in {2,6,10,14,18,22,26,30}
                \draw[thick, cyan] ({\i},0) -- ({\i+2},0);
        \foreach \i in {2,4,6,8,10,12,14,16,18,20,22,24,26,28,30,32}
                \filldraw[black] ({\i},0) circle (0.15);
        \end{tikzpicture}
        \caption{toric code with no defect}\label{fig: no defect}
    \end{subfigure} 

    \begin{subfigure}{\textwidth}
        \centering
        \begin{tikzpicture}[scale=0.27]
        \foreach \i in {2,6,10,14,18,22,26,30}
                \draw[thick, black] ({\i},0) -- ({\i},3);
        \foreach \i in {2,6,10,14,18,22,26,30}
                \draw[thick, black] ({\i+2},0) -- ({\i+2},-3);
        \foreach \i in {2,8,14,18,22,26,30,34}
                \draw[thick, yellow] ({\i},0) -- ({\i-2},0);
        \foreach \i in {2,4,8,10,14,18,22,26,30}
                \draw[thick, cyan] ({\i},0) -- ({\i+2},0);
        \foreach \i in {2,4,6,8,10,12,14,16,18,20,22,24,26,28,30,32}
                \filldraw[black] ({\i},0) circle (0.15);
        \draw[red] (4,0) circle (0.3);
        \draw[red] (10,0) circle (0.3);
        \end{tikzpicture}
        \caption{toric code with a pair of defects labelled by red circles}\label{fig:2 defects}
    \end{subfigure} 

    \begin{subfigure}{\textwidth}
        \centering
        \begin{tikzpicture}[scale=0.27]
        \foreach \i in {2,6,10,14,18,22,26,30}
                \draw[thick, black] ({\i},0) -- ({\i},3);
        \foreach \i in {2,6,10,14,18,22,26,30}
                \draw[thick, black] ({\i+2},0) -- ({\i+2},-3);
        \foreach \i in {2,8,12,18,22,26,30,34}
                \draw[thick, yellow] ({\i},0) -- ({\i-2},0);
        \foreach \i in {2,4,8,12,14,18,22,26,30}
                \draw[thick, cyan] ({\i},0) -- ({\i+2},0);
        \foreach \i in {2,4,6,8,10,12,14,16,18,20,22,24,26,28,30,32}
                \filldraw[black] ({\i},0) circle (0.15);
        \draw[red] (4,0) circle (0.3);
        \draw[red] (14,0) circle (0.3);
        \end{tikzpicture}
        \caption{Illustrates The right defect being moved one step to the right.}\label{subfig: longer defect}
    \end{subfigure} 

    \begin{subfigure}{\textwidth}
        \centering
        \begin{tikzpicture}[scale=0.27]
        \foreach \i in {2,6,10,14,18,22,26,30}
                \draw[thick, black] ({\i},0) -- ({\i},3);
        \foreach \i in {2,6,10,14,18,22,26,30}
                \draw[thick, black] ({\i+2},0) -- ({\i+2},-3);
        \foreach \i in {2,8,12,18,24,30,34}
                \draw[thick, yellow] ({\i},0) -- ({\i-2},0);
        \foreach \i in {2,4,8,12,14,18,20,24,26,30}
                \draw[thick, cyan] ({\i},0) -- ({\i+2},0);
        \foreach \i in {2,4,6,8,10,12,14,16,18,20,22,24,26,28,30,32}
                \filldraw[black] ({\i},0) circle (0.15);
        \draw[red] (4,0) circle (0.3);
        \draw[red] (14,0) circle (0.3);
        \draw[red] (20,0) circle (0.3);
        \draw[red] (26,0) circle (0.3);
        \end{tikzpicture}
        \caption{Toric code with two pairs of defects}\label{fig:4 defects}
    \end{subfigure}

    \begin{subfigure}{\textwidth}
        \centering
        \begin{tikzpicture}[scale=0.27]
        \foreach \i in {2,6,10,14,18,22,26,30}
                \draw[thick, black] ({\i},0) -- ({\i},3);
        \foreach \i in {2,6,10,14,18,22,26,30}
                \draw[thick, black] ({\i+2},0) -- ({\i+2},-3);
        \foreach \i in {2,8,12,16,20,24,30,34}
                \draw[thick, yellow] ({\i},0) -- ({\i-2},0);
        \foreach \i in {2,4,8,12,16,20,24,26,30}
                \draw[thick, cyan] ({\i},0) -- ({\i+2},0);
        \foreach \i in {2,4,6,8,10,12,14,16,18,20,22,24,26,28,30,32}
                \filldraw[black] ({\i},0) circle (0.15);
        \draw[red] (4,0) circle (0.3);
        \draw[red] (26,0) circle (0.3);
        \end{tikzpicture}
        \caption{ Illustrates the process of fusing the two middle defects, which merges the separate pairs into a single pair.}\label{fig:fused}
    \end{subfigure} 
    
    \caption{The figures presented above depict the static Hamiltonian on a portion of the lattice. It is required that all check operators on yellow edges be dominant. Different choices of dominant check operators will lead to various cases of the effective toric code Hamiltonian, with or without defects.}
    \label{fig:static hamiltonian}
\end{figure}

The defect remains at the trivalent vertex, as a degree of three disrupts the local 2-colorability, as indicated in Figure~\ref{fig:lattice with a defect}. Therefore, moving the defect involves relocating the trivalent vertex.

Consider Figure~\ref{fig:static hamiltonian}, which depicts a section of a larger lattice, similar to those shown in Figures~\ref{fig:lattice without defect} and~\ref{fig:lattice with defect}. All check operators on yellow edges are designated as dominant. The configuration outside this localized area remains unchanged. This setup presents five potential configurations, where the effective Hamiltonian can represent the toric code, with or without defects. To transition between these static states, we introduce time evolution, facilitating the creation, movement, and fusion of defects.

\begin{figure}[htbp]
    \centering
    \begin{subfigure}{0.4\textwidth}
        \centering
        \begin{tikzpicture}[scale=0.3]
        \draw[thick, blue] (-8,0) -- (8,0);
        \draw[thick, blue] (0,0) -- (-4,-5);
        \foreach \i in {-4,4}
                \draw[thick, blue] ({\i},0) -- ({\i},4);
        \foreach \i in {-4,0,4}
                \filldraw[black] ({\i},0) circle (0.2);
        \node[black] at (-8.5,0.5) {\tiny $X$};
        \node[black] at (-5,0.5) {\tiny $X$};
        \node[black] at (1,0.5) {\tiny $X$};
        \node[black] at (3,0.5) {\tiny $X$};
        \node[black] at (-4.5,1.5) {\tiny $Y$};
        \node[black] at (-4.5,3.5) {\tiny $Y$};
        \node[black] at (4.5,1.5) {\tiny $Y$};
        \node[black] at (4.5,3.5) {\tiny $Y$};
        \node[black] at (-0.3,-1.2) {\tiny $Y$};
        \node[black] at (-2.8,-4.4) {\tiny $Y$};
        \node[black] at (-3,0.5) {\tiny $Z$};
        \node[black] at (-1,0.5) {\tiny $Z$};
        \node[black] at (5,0.5) {\tiny $Z$};
        \node[black] at (8.5,0.5) {\tiny $Z$};
        \end{tikzpicture}
        \caption{}
        \label{time evolusion lattice detail}
    \end{subfigure}
    \hfill
    \begin{subfigure}{0.4\textwidth}
        \centering
        \includegraphics[width=\linewidth]{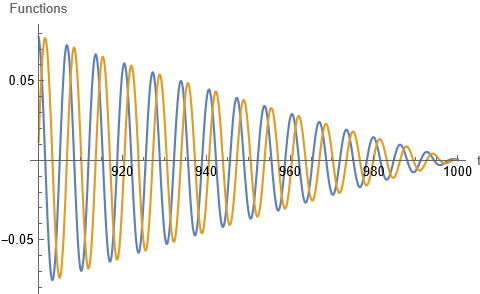}
        \caption{}
        \label{numerical suggestion}
    \end{subfigure}
    \hfill

    \caption{The figure on the left depicts a detailed local part of a honeycomb lattice to elucidate the movement of a defect, with the check operator explicitly labeled. On the right, the numerical results are displayed, illustrating that the real and imaginary differences between $\beta_1$ and $\beta_2$ vanish at $T = 1000$. Furthermore, this pattern persists for all $T > 1000$.}
    \label{fig:detalied moving process}
\end{figure}
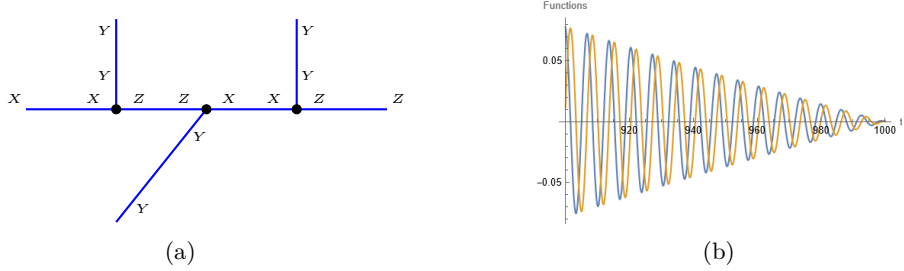

Focusing on the movement of a defect as a detailed example (the other processes are analogous), we examine a more specific local structure in Figure~\ref{fig:detalied moving process}, which illustrates the transformation from Figure~\ref{fig:2 defects} to Figure~\ref{subfig: longer defect}. We use $H(0)$ and $H(T)$ to denote their respective static Hamiltonians. The linear interpolation between them is introduced as follows:
\begin{equation}
    H(t) = H(0) \left(1 - \frac{t}{T}\right) + H(T) \frac{t}{T}
\end{equation}
$H(0)$ and $H(T)$ represents the Hamiltonian with dominant coefficients of $S_c(0)$ and $S_c(T)$ and all perturbation terms were shut down to avoid subtlety. $H(t)$ commutes with all the plaquette terms so the action of all plaquette terms remains unchanged. Therefore, the action of the time evolution operator on the stabilizer center$S_c$ is crucial. We expect the state will transition into the spectrum of new stabilizer centers.

Note that most terms remain unchanged, contributing to a constant phase, as the state is always their eigenstate with an eigenvalue of $+1$. The only non-trivial terms are: 
\begin{equation}
    H = -\frac{t}{T} Z\otimes Z \otimes 1 +- (1-\frac{t}{T}) 1 \otimes X \otimes X 
\end{equation}
The time evolution operator(TEO):
\begin{equation}\label{eqn:time evolution operator}
    O(t)= \mathcal{T} \int e^{i H}dt
\end{equation}
This is a formal notation; calculations need to be done by explicitly applying the Time Order operator $\mathcal{T}$. But we realize that after expanding $O(t)$, the general form is:
\begin{equation}
    O(t)=a(t)+b(t) Z\otimes Z \otimes 1 +c(t) 1 \otimes X \otimes X  +d(t) Z \otimes Y \otimes X
\end{equation}
Where $a, b, c,$ and $d$ are complex, time-dependent functions. Since the operator $O(t)$ acts on the ground state of $1 \otimes X \otimes X$,  a simplified representation is allowed due to the trivial action of $1 \otimes X \otimes X$:
\begin{equation}
    O(t) = \beta_1 + \beta_2 Z \otimes Z \otimes 1
\end{equation}
Utilizing the differential equation
\begin{equation}
    \frac{d O(t)}{dt} = H O(t)
\end{equation}
we numerically solve for $O(t)$, finding that at time $T$, $O(T) = \beta (1 + Z \otimes Z \otimes 1)$. Here, $\beta$ is a complex number whose significance is determined by the value of $T$. Assuming the ground state initiates as
\begin{equation}\label{eqn:t=0 ground state}
    |GS\rangle_{t=0} = \Pi_{p \in P} \frac{W_p + 1}{2} \bigotimes_{v' \in V'} |0\rangle_{v'}
\end{equation}
Remember $V'$ is the set of vertices on the shrunken lattice. Since the plaquette operators commute with the time evolution operator, the ground state transitions to:

\begin{equation} \label{eqn:t=T time evoluted state}
    |GS\rangle_{t=T} =  \Pi_{p\in P} \frac{W_p+1}{2} \left\{ O(T) \bigotimes_{v\in V'} |0\rangle \right\}
\end{equation}
This is equivalent to the ground state of $H(T)$! Thus we claim we are able to move the defect.

%At $t=0$, the three qubits have two effective qubits with following basis.
%\begin{align} \label{eqn:t=0 basis}
%    |00\rangle_{t=0} &= |0\rangle (|00\rangle+|01\rangle+|10\rangle+|11\rangle)\\
%    |01\rangle_{t=0} &= |0\rangle (|00\rangle-|01\rangle-|10\rangle+|11\rangle)\\
%    |10\rangle_{t=0} &= |1\rangle (|00\rangle+|01\rangle+|10\rangle+|11\rangle)\\
%    |00\rangle_{t=0} &= |1\rangle (-|00\rangle+|01\rangle+|10\rangle-|11\rangle)\\ 
%\end{align}
%The basis chosen is to keep the action of the surrounding unperturbed plaquette terms unchanged. Since the ground state remains:

%Where $V'$ is the vertex set of the shrunk lattice. And the $|0\rangle_{v'}$ refer to the effective qubit state. It should be noted that equation \ref{eqn:t=0 basis} shows that the two effective qubits which are closest to the defect are slightly entangled. This is attributed to the left pentagon operator, located below and caused by the defect, that slightly entangles them. We then apply the $O(t=T)$ to get new basis:
%\begin{align} \label{eqn:t=T basis}
%    |00\rangle_{t=T} &= |00\rangle (|0\rangle+|1\rangle)\\
%    |01\rangle_{t=T} &= |00\rangle (|0\rangle-|1\rangle)\\
%    |10\rangle_{t=T} &= |00\rangle (|0\rangle+|1\rangle)\\
%    |11\rangle_{t=T} &= |00\rangle (|0\rangle-|1\rangle)\\
%\end{align}
%Notice the aforementioned entanglement is  removed since the defect is moved away.

    \begin{figure}[htbp]
\centering
\begin{tikzpicture}[scale=0.4]
\foreach \i in {2,6,10,14}
        \draw[thick, black] ({\i},0) -- ({\i},2);
\foreach \i in {4,8,12,16}
        \draw[thick, black] ({\i},0) -- ({\i},-3);
\foreach \i in {2,6,10,14,18}
        \draw[thick, yellow] ({\i},0) -- ({\i-2},0);
\foreach \i in {2,6,10,14}
        \draw[thick, cyan] ({\i},0) -- ({\i+2},0);
\foreach \i in {0,2,4,6,8,10,12,14,16,18}
        \filldraw[black] ({\i},0) circle (0.15);
\foreach \i in {0,6,12,18}
        \draw[red] ({\i},0) circle (0.3);
\foreach \i in {0,6,12,18}
        \draw[red] ({\i},0) circle (0.3);
\foreach \i in {1,3,5,7,9,11,13,15,17}
        \draw[gray] ({\i},-0.8) circle (0.4);
\node[gray] at (1,-0.8) {\scriptsize $1$};
\node[gray] at (3,-0.8) {\scriptsize $2$};
\node[gray] at (5,-0.8) {\scriptsize $3$};
\node[gray] at (7,-0.8) {\scriptsize $4$};
\node[gray] at (9,-0.8) {\scriptsize $5$};
\node[gray] at (11,-0.8) {\scriptsize $6$};
\node[gray] at (13,-0.8) {\scriptsize $7$};
\node[gray] at (15,-0.8) {\scriptsize $8$};
\node[gray] at (17,-0.8) {\scriptsize $9$};
\end{tikzpicture}
\caption{The image displays a local section of the lattice, where alternating red and blue edges represent the $X \otimes X$ and $Z \otimes Z$ checks respectively. Each edge is labeled by a unique number, with the check operator of the corresponding edge identified accordingly. The red circles indicate the positions intended for defect placement.}
\label{fig:illustration of fusion process}
\end{figure}
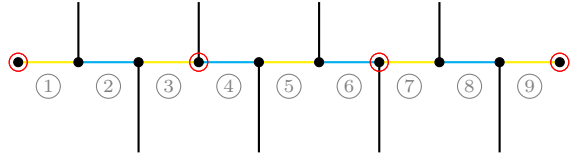

It is instructive to concentrate on a single chain as depicted in Figure~\ref{fig:illustration of fusion process}. In this illustration, red edges correspond to the check operator $X \otimes X$, and blue edges to $Z \otimes Z$. These operators are denoted by $Op_i$, where $i$ signifies the numerical assignment to the edges. Initially, all blue checks are designated as part of the Stabilizer center.

Within this framework, we explore several critical processes. The initial process entails the creation of defects via the application of the Hamiltonian:
\begin{equation}
    H(t) = -(Op_1 + Op_3) \left(1 - \frac{t}{T}\right) - (Op_2) \frac{t}{T}
\end{equation}
Terms that commute with $H(t)$ are omitted. A numerical solution reveals that the time evolution operator is expressed as:
\begin{equation}
    O^{creation}(T) = \beta(T) (1 + Op_2)
\end{equation}
This implies that the state will be projected onto the ground state of $Op_2$, as anticipated, leading to the creation of a pair of defects. This transformation is represented in the transition from Figure~\ref{fig: no defect} to Figure~\ref{fig:2 defects}.

In the second scenario, the movement of one of the defects is achieved through the following process:
\begin{equation}
    H(t) = -(Op_2 + Op_5) \left(1 - \frac{t}{T}\right) - (Op_2 + Op_4) \frac{t}{T}
\end{equation}
As explicitly demonstrated above, this time evolution is as expected to move defects.

The final process to consider is the fusion of defects. The initial case of fusion involves creating a pair of defects and subsequently fusing them back together, essentially reversing the creation process:
\begin{equation}
    H(t) = -Op_2 \left(1 - \frac{t}{T}\right) - (Op_1 + Op_3) \frac{t}{T}
\end{equation}
The numerical solution for the time evolution operator is given by:
\begin{flalign}\label{eqn:TEO of fusion process}
    O(T) =\ &\beta_3(T) (1 + Op_1)(1 + Op_3) &\nonumber \\
    &+ \beta_4(T) \left[(1 - Op_1)(1 + Op_3) + (1 + Op_1)(1 - Op_3)\right]
\end{flalign}

In this scenario, only $(1 + Op_1)(1 + Op_3)$ has non-zero action by examining energy levels. This indicates that upon fusing the pair created from the vacuum, we algebraically regain the vacuum state as expected. A more intriguing case of fusion involves creating two pairs of defects from the vacuum, as depicted in Figure~\ref{fig:4 defects}, and then fusing the two central defects. Denote the state before fusion as $|GS_4\rangle$, derived from creating four defects from $|GS\rangle$, as seen in Equation~\ref{eqn:t=0 ground state}. The process then transitions these two pairs into a single pair:

\begin{equation}
 H(t)=-Op_5(1-\frac{t}{T})-(Op_4+Op_6)\frac{t}{T}
\end{equation}
The TEO is similar as \ref{eqn:TEO of fusion process}:
\begin{equation}
    O(T)=\beta_5(T) (1+Op_4)(1+Op_6)+\beta_6(T) [(1-Op_4)(1+Op_6)+(1+Op_4)(1-Op_6)]
\end{equation}\label{eqn:TEO of fusion process 2}
To check the fusion rule of the defects. We should check the normalization of the projectors. We will see $\langle GS_4|(1 \pm Op_4)(1 \pm Op_6)|GS_4\rangle$ is consistently identical. To understand this, notice that:
\begin{equation}
    \langle GS_4|Op_{4}|GS_4\rangle=\langle GS_4|Op_{6}|GS_4\rangle= 0
\end{equation}
This is because  $Op_{4}(Op_6)|GS_4\rangle$ has different energy from $|GS_4\rangle$, as $\frac{1+Op_5}{2} |GS_4\rangle=|GS_4\rangle$ and $Op_4\frac{1+Op_5}{2} |GS_4\rangle=\frac{1-Op_5}{2}Op_4|GS_4\rangle$. The intricate part is:
\begin{equation}
    \langle GS_4|Op_4 \otimes Op_6|GS_4\rangle= \left( \bigotimes_{v' \in V'}\langle 0|_{v'} \right)
    \Pi_{p\in P} \frac{W_p+1}{2} Op_4 \otimes Op_6 \left( \bigotimes_{v''\in V'} |0\rangle_{v''} \right)
\end{equation}
Remember $V'$ here represents the set of vertices of the shrunken lattice. Notice:
\begin{equation}
    \left( \bigotimes_{v' \in V'}\langle 0|_{v'} \right)
    \Pi_{p\in P'} W_p Op_4 \otimes Op_6 \left( \bigotimes_{v''\in V'} |0\rangle_{v''} \right)=0
\end{equation}
Because, for any $P' \subset P$, the product of $W_p$s always acts on a trivial loop of the lattice, which can not match the action of the  $Op_4 \otimes Op_6$ on an open cut. And $Op_4 \otimes Op_6$ itself can not act trivially on $\bigotimes_{v\in V'} |0\rangle_{v}$.

Naturally, $(1+Op_4)(1+Op_6)$ yields the vacuum, while $(1-Op_4)(1+Op_6)+(1+Op_4)(1-Op_6)$ gives rise to a free fermion. If fusion rule obeys the rule of the Ising Anyon, their coefficients should satisfy:
\begin{equation}
    \beta_5(T)=\sqrt{2} \beta_6(T)
\end{equation}
Numerical solutions suggest that $|\beta_5(T)|=\sqrt{2} |\beta_6(T)|$, with a surprisingly introduced phase. However, we can account for this by moving the phase into the definition of the state or choose the $T$ carefully to let the phase vanish.
As highlighted in \cite{Kitaev_2006}, the free fermion excitation exhibits the same algebra as the composite quasi-particle of electric and magnetic charge $\epsilon$ of the toric code, even though they differ in energy. Kitaev proposed that the free fermion would decay to $\epsilon$ when exposed to a certain thermal bath. Consequently, we can deduce that the defects explicitly comply with the nontrivial fusion rule of the Ising Anyon as demonstrated in \cite{Bombin_2010}:
\begin{equation}
    \sigma \times \sigma = 1 + \epsilon
\end{equation}
$\sigma$ represents the twist defect. $\epsilon$ represents the fermion.
The last thing we have to care is that although the action of plaquette terms are fixed during time evolution, the effective Hamiltonian may flip its sign so it may have excitations which violates plaquette terms.

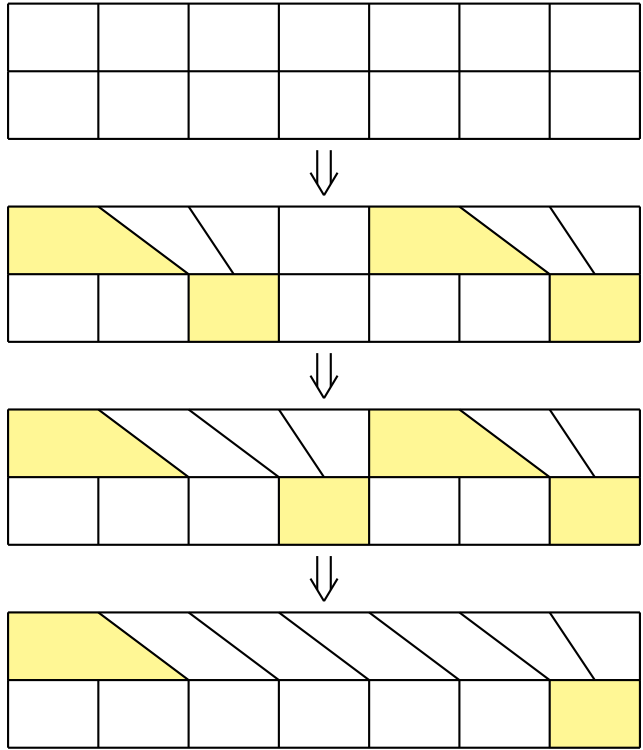
\begin{figure}[ht]
\centering
\begin{tikzpicture}[scale=0.3]
\foreach \j in {-10.5, -1.5, 7.5}
        \draw[black, fill=yellow!80, opacity=0.6] (0,{\j}) -- (4,{\j}) -- (8,{\j-3}) -- (0,{\j-3}) -- cycle;
\foreach \j in {-1.5, 7.5}
        \draw[black, fill=yellow!80, opacity=0.6] (16,{\j}) -- (20,{\j}) -- (24,{\j-3}) -- (16,{\j-3}) -- cycle;
\foreach \j in {-13.5, -4.5, 4.5}
        \draw[black, fill=yellow!80, opacity=0.6] (24,{\j}) -- (28,{\j}) -- (28,{\j-3}) -- (24,{\j-3}) -- cycle;
\draw[black, fill=yellow!80, opacity=0.6] (8,4.5) -- (12,4.5) -- (12,1.5) -- (8,1.5) -- cycle;
\draw[black, fill=yellow!80, opacity=0.6] (12,-4.5) -- (16,-4.5) -- (16,-7.5) -- (12,-7.5) -- cycle;

\foreach \j in {-16.5,-13.5,-10.5,-7.5,-4.5,-1.5,1.5,4.5,7.5,10.5,13.5,16.5}
        \draw[thick, black] (0,{\j}) -- (28,{\j});        
\foreach \i in {0, 4, 8, 12, 16, 20, 24, 28}
\foreach \j in {-16.5, -7.5, 1.5, 10.5, 13.5}
        \draw[thick, black] ({\i},{\j}) -- ({\i},{\j+3});

\foreach \j in {-13.5, -4.5, 4.5}
        \draw[thick, black] (0,{\j}) -- (0,{\j+3});
\foreach \j in {-13.5, -4.5, 4.5}
        \draw[thick, black] (28,{\j}) -- (28,{\j+3});
\foreach \j in {-13.5, -4.5, 4.5}
        \draw[thick, black] (26,{\j}) -- (24,{\j+3});
\foreach \j in {-4.5, 4.5}
        \draw[thick, black] (24,{\j}) -- (20,{\j+3});
\foreach \j in {-4.5, 4.5}
        \draw[thick, black] (16,{\j}) -- (16,{\j+3});
\foreach \j in {-4.5, 4.5}
        \draw[thick, black] (8,{\j}) -- (4,{\j+3});
\foreach \i in {4, 8, 12, 16, 20}
        \draw[thick, black] ({\i},-10.5) -- ({\i+4},-13.5);
\draw[thick, black] (12,4.5) -- (12,7.5);
\draw[thick, black] (10,4.5) -- (8,7.5);
\draw[thick, black] (14,-4.5) -- (12,-1.5);
\draw[thick, black] (12,-4.5) -- (8,-1.5);

\foreach \i in {13.7, 14.3}
\foreach \j in {-8, 1, 10}
        \draw[thick, black] ({\i},{\j}) -- ({\i},{\j-1.5});
\foreach \j in {-10, -1, 8}
        \draw[thick, black] (14,{\j}) -- (14.6,{\j+1});
\foreach \j in {-10, -1, 8}
        \draw[thick, black] (14,{\j}) -- (13.4,{\j+1});
\end{tikzpicture}
\caption{Demonstration of the excited plaquettes. Yellow plaquettes are the excited ones while white plaquettes stays at the ground state of the corresponding plaquette operators. The first figure is a part of regular surface code case and the state stays at the ground state. When two pairs of defects are created, two pairs of plaquettes, which have the trivalent vertex, will be excited. Moving defects will move the excited plaquettes accordingly with no more excitation created. After fusing the central two defects, the corresponding excitation annihilates, leaving two remaining defects at ends.}
\label{fig:excited plaquettes}
\end{figure}

To see this, rewrite the overall Hamiltonian in a simplified manner:
\begin{equation}
    H = -S_c- c_p W_p
\end{equation}
$c_p$ absorbs all coefficients of the plaquette operators and $S_c$ are dominant.  
From explicit numerical evaluation in appendix(\ref{appendx:effective Hamiltonian}), the sign of two plaquette operator flips after creating or fusing a pair of defects. So two extra excitations appear or annihilate but moving defects won't create any excitation as shown in diagram\ref{fig:excited plaquettes}.  The yellow plaquette is the one that is excited while white plaquette stays at the ground state of the corresponding $W_p$. The overall picture is: create two pairs of defects and four plaquettes carry plaquette excitations. Then the central two defects fused with two plaquette excitation annihilates, leaving a superposition of vacuum and free fermion excitation, which agrees with the picture that the defects capture the Majorana fermion and behaves like Ising anyon.

\section{Subsystem Code aspects} \label{sec:subsystem code}
In recent work of \cite{Vuillot_2021}, the Kitaev spin liquid code on trivalent and 3-colorable lattice has been proved to be a zero-logical-qubit 
subsystem code. Here we generalize it to our case. We use some notations to describe the lattice by $n_v$, the number of vertices, $n_e$, the number of edges, $n_p$, the number of plaquettes.
We only look at orientable lattice, which could be easily extended to non-orientable cases. We would first prove it is true when all the vertices of lattice have odd degrees. The gauge group is generated by $P_e|e\in E$. So the number of gauge group generators is $n_e-1$ due to the $\Pi_{e\in E} P_e=1$. Then the generators of stabilizer group $S$ is generated by $\{W_p | p\in P \} \cup \{W_l\}$, where ${W_l}$ denotes the set of operators that are formed by the product of check operators along non-trivial loops of the lattice. So the number of the generators of $S$ is given by $n_p-1+k$, where $k$ is the number of non-trivial loop on the lattice. Assume vertex $v$ has degree $d_v$ and $t_v$ qubits are placed on it. Then:
\begin{align}
    n_v-n_e+n_p & = 2-k\\
    2 n_e & = \sum_{v} d_v \\
    n_q&=\sum_{v} t_v
\end{align}
Where $n_q$ is the number of total qubits. The number of logical qubit $n_L$ of this subsystem code is given by
\begin{align}
    n_L & =  n_q-(n_g-n_s)/2-n_s\\
    & =  1/2 \, (2 n_q-n_g-n_s)\\
    & =  1/2 \, (2 n_q-n_e-n_p+2-k)\\
    & =  1/2 \,(2 n_1-n_e+n_v-n_e)\\
    & =  1/2 \, (\sum_{v\in V} {2 t_v+1-d_v})
\end{align}
In our setup, the number of qubit on an odd degree vertex is given by $t_v = (d_v-1)/2 $. So it will automatically let $n_L=0$. 
An even degree vertex could be treated as two connected odd degree vertices as depicted in Fig. \ref{fig:2 odd to one even vertex}, but in a converse manner. Clearly, this splitting does not change any of the aforementioned total quantities. So our generalized two dimensional Kitaev spin liquid model is always a zero-logical-qubit subsystem. The implication of floquet code is possible but out of the scope of this paper.

\section{Conclusion and outlook} \label{sec:conclusion and outlook}
In this paper, we have generalized the Kitaev spin liquid model on a general planar lattice. We proposed that if we can identify a stabilizer center $S_c$ to satisfy certain requirement, that $S_c$ contains maximum amount of commuting  check operators, the vicinity of the ground state of $S_c$ will be effectively toric code model. If a single trivalent vertex remains in the shrunken lattice, a pair of twist defects would emerge, exhibiting Non-Abelian statistics as Ising Anyons. We have conclusively shown that we can manipulate and fuse the defect as long as the Hamiltonian is altered slowly. Furthermore, the processes of creation, movement, and fusion are all achieved by the time evolution operator, which are inherently unitary operators. It is equivalent to say we can use unitary operators to create, move and fuse defects, which aligns with our usual taste of manipulating anyons. Nonetheless, braiding continues to pose a challenge in this context. In conclusion, the generalized spin liquid model appears to be a versatile platform for realizing a general surface code. 

Several promising directions for future research emerge from this study. For instance, the nature of a defect resulting from a left vertex of degree 5 remains to be explored. The algebra looks similar but it creates defect disrupting more plaquettes. An extension to describe three dimensional topological phases or Fractonic phases would also be an intriguing prospect, and is currently under preparation. An analytical calculation of the geometric factor $\alpha_p$ may be interesting since the numerical calculation yields highly regular and interesting results. Moreover, a more general and analogous generalization that could support the non-Abelian Kitaev Quantum Double model would be of significant interest and importance.

\vspace{0.2cm}
\noindent\textbf{Acknowledgments.}

The authors are partially supported by NSF CCF 2006667, Quantum Science Center (led by ORNL), and ARO MURI.

\printbibliography

\appendix
\section{Perturbation treatment} \label{appendix:perturbation theory}
\subsection{Effective Hamiltonian}\label{appendx:effective Hamiltonian}
In this appendix, $V$ represents the perturbation $H^{'}$ to avoid confusion. For the effective Hamiltonian as follows:
\begin{align}
\centering
    H_eff = T (V+V G_0^{'}  V+V G_0^{'} V G_0^{'} V+\ldots)T
\end{align}
Where, $T = |GS\rangle \langle GS|$ is the projector onto the ground state of $J_v$s, and $G_0^{'}=(1/(E_0-H_0))'$, where the prime notation implies $G_0^{'}$ vanishes on the ground state and acts normally on the excited states.

\begin{figure}
\centering
\begin{tikzpicture}[
  node distance=2cm and 2cm,
  block/.style={draw, rectangle, rounded corners},
  arrow/.style={thick,->,>=stealth},
]

% Column 1
\node [block] (A) {$|\Phi_0\rangle,0$};

% Column 2
\node [block, right=of A] (B) {$|\Phi_1\rangle,4$};

% Column 3
\node [block, above right=of B,yshift=+1.5cm] (C1) {$|\Phi_{21}\rangle,0$};
\node [block, right=of B] (C2) {$|\Phi_{22}\rangle,4$};
\node [block, below right=of B,yshift=-1.5cm] (C3) {$|\Phi_{23}\rangle,8$};

% Column 4

\node [block, right=of C1] (D2) {$|0\rangle,0$};

\node [block, above right=of C2, yshift=-1cm] (D4) {$|\Phi_{31}\rangle,0$};
\node [block, right=of C2] (D5) {$|\Phi_{32}\rangle,4$};
\node [block, below right=of C2, yshift=+1cm] (D6) {$|\Phi_{33}\rangle,8$};

\node [block, above right=of C3, yshift=-1cm] (D7) {$|\Phi_{34}\rangle,4$};
\node [block, right=of C3] (D8) {$|\Phi_{35}\rangle,8$};
\node [block, below right=of C3, yshift=+1cm] (D9) {$|\Phi_{36}\rangle,12$};

% Arrows
\draw [arrow] (A) -- (B) node[midway,above] {$V$};

\draw [arrow] (B) -- (C1)node[midway,above left] {$V G_0^{'}$};
\draw [arrow] (B) -- (C2)node[midway,above ] {$V G_0^{'}$};
\draw [arrow] (B) -- (C3)node[midway, left] {$V G_0^{'}$};

\draw [arrow] (C2) -- (D4)node[midway,above left] {$V G_0^{'}$};
\draw [arrow] (C2) -- (D5)node[midway,above ] {$V G_0^{'}$};
\draw [arrow] (C2) -- (D6)node[midway, left] {$V G_0^{'}$};

\draw [arrow] (C3) -- (D7)node[midway,above left] {$V G_0^{'}$};
\draw [arrow] (C3) -- (D8)node[midway,above ] {$V G_0^{'}$};
\draw [arrow] (C3) -- (D9)node[midway, left] {$V G_0^{'}$};

\draw [arrow] (C1) -- (D2) node [midway,above ] {$VG_0^{'}$};

\end{tikzpicture}
\caption{Demostration of perturbation tree}
\label{fig:perturbation}
\end{figure}
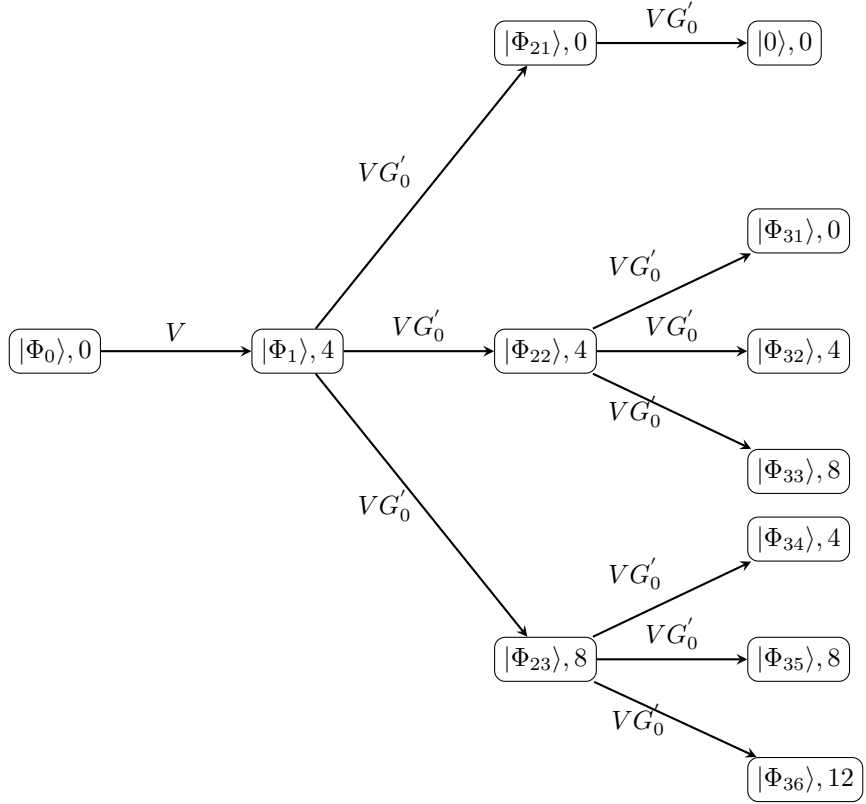

The perturbation tree, shown in figure\ref{fig:perturbation}, demonstrates the general idea of calculation. Each node of the diagram is labeled with a state $\Phi$. The action of each check operator on $|\Phi_0\rangle = |GS\rangle$ will flip two terms in $H_0$, which increases the energy by 4. Then, when we apply an another term in $V$ on the intermediate state $|\Phi_1\rangle$, different outcomes appear due to the product of $P_{e_1} \times P_{e_2}$, where:

If $\partial e_1 \cap \partial e_2 = \emptyset$, then we obtain $|\Phi_{23}\rangle$ with energy 8.
If $\partial e_1 = \partial e_2$, then we obtain $|\Phi_{21}\rangle$, which goes back to the ground state.
The remaining possibility is that $e_1$ and $e_2$ share one common vertex and will have energy 4.
If we repeat this process to achieve higher order perturbation, we obtain different sectors with different energy levels as shown in Figure \ref{fig:perturbation}. The energy sector $\Phi_{i,j}$ that reverts back to the ground state will contribute to the $i$th order of the effective Hamiltonian. Note that in the diagram, $G_0^{'}$ always gives some constant factor. Thus, we conclude:
\begin{equation}
    H_{eff}^i=\alpha_i \langle GS|V^i|GS\rangle
\end{equation}

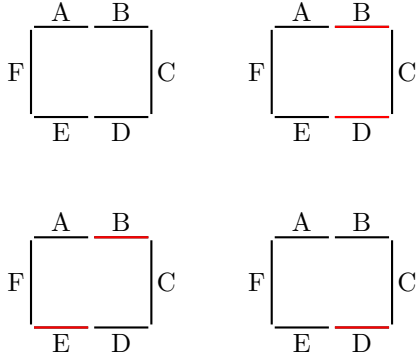
\begin{figure}[ht]
\centering
\begin{tikzpicture}[scale=0.4]
\foreach \i in {-6, -2, 2, 6}
\foreach \j in {2, -5}
        \draw[thick, black] ({\i},{\j+0.1}) -- ({\i},{\j+2.9});
\foreach \i in {-6, -4, 2, 4}
\foreach \j in {-5, -2, 2, 5}
        \draw[thick, black] ({\i+0.1},{\j}) -- ({\i+1.9},{\j});
\foreach \i in {-5, 3}
\foreach \j in {-5.5, 1.5}
        \node[black] at ({\i},{\j}) {E};
\foreach \i in {-3, 5}
\foreach \j in {-5.5, 1.5}
        \node[black] at ({\i},{\j}) {D};
\foreach \i in {-5, 3}
\foreach \j in {-1.5, 5.5}
        \node[black] at ({\i},{\j}) {A};
\foreach \i in {-3, 5}
\foreach \j in {-1.5, 5.5}
        \node[black] at ({\i},{\j}) {B};
\foreach \i in {-6.5, 1.5}
\foreach \j in {-3.5, 3.5}
        \node[black] at ({\i},{\j}) {F};
\foreach \i in {-1.5, 6.5}
\foreach \j in {-3.5, 3.5}
        \node[black] at ({\i},{\j}) {C};
\foreach \j in {-5, 2, 5}
        \draw[thick, red] (4.1,{\j}) -- (5.9,{\j});
\draw[thick, red] (-2.1,-2) -- (-3.9,-2);
\draw[thick, red] (-4.1,-5) -- (-5.9,-5);
\end{tikzpicture}
\caption{The figures above display a plaquette with $6$ edges, where the red edges represent the dominant check operators and the black edges denote the perturbation terms. The figure in the top-right position illustrates a common scenario in the honeycomb lattice. The figure in the bottom-left position shows the case after moving defects, and the figure in the bottom-right position depicts the situation when there is a defect}
\label{fig:explanation of the sign of W_p}
\end{figure}

Where $\alpha_i$ is a path-dependent factor calculable from the diagram. We name it the geometric factor. Consider the order of $\lambda$ first.  Terms in $V^i$ that do not vanish in the ground state are those that commute with all $P_v$. Generators could be $P_{e_i} P_{e_j}$, where $e_i = e_j$, contributing to a constant factor, or $\Pi_{e_j} P_{e_j}$, where $e_j$ forms a closed loop, contributing to the plaquette term $W_p^{'} = \Pi_{e\in Bo(p)|P_e \in H^{'}} P_e$. 
So we could conclude, the effective Hamiltonian is:
\begin{equation}
    H_{eff}= \sum_p (-1)^{\gamma_p}\alpha_{l_p} \lambda^{l_p} W_p +  constant
\end{equation}
Where $l_p$ is the number of edges that appears in $W_p^{'}$. The factor $\lambda^{l_p}$ arises from the dominant contribution in the $l_p$th order perturbation. 
We have to pay extra attention to the plaquette operators $W_p=\Pi_{e\in Bo(p)} P_e$. But the perturbation will end in $W_p^{'}$, an incomplete form of the $W_p$. But we can recover $W_p$ by adding the dominant check operators.
To understand this, look at a special case as figure\ref{fig:explanation of the sign of W_p}. A plaquette consists of $6$ edges and the check operators attached to them are labeled by $A,B,C,D,E,F$. Consider $B,D$ is dominant as shown in the right above case, then $A,C,E,F$ are perturbation terms. Numerical calculation shows that the perturbation gives $W_p^{'}= \alpha_p ACEF = -\frac{1}{16} ACEF$. We can easily recover $W_p$ by adding $B$ and $D$ by $W_p^{'} B D=-\frac{1}{16} ACEF B D=-\frac{1}{16} ABCDEF$, since the state is the ground state of $B$ and $D$ with eigenvalue $+1$. This means the effective Hamiltonian can be written as $H_{eff}=-\frac{1}{16} W_p$. By this, the effective Hamiltonian could be fanalized to form\ref{eqn: effective Hamiltonian formula}.

After the process of moving defects, it becomes the case that $B$ and $E$ are dominant, and we will get the exact same effective Hamiltonian by the same way. Since the action $W_p$ remains invariant during time evolution. Combined with the fact that the state is the ground state of the dominant $S_c(T)$, we conclude it will ends in the ground state of the $H(T)$.

The cases of creation or fusion cases are different. In these cases, the plaquette will be transformed into cases(reversely for fusion process) that only $D$ is dominant. The perturbation method gives $-\frac{1}{16} ABCEF $. So $H_{eff}=-\frac{1}{16} ABCEF D=\frac{1}{16}ABCDEF$. Thus, though $W_p$ remains the same eigenvalue, it ends in excitation state since the different sign of the plaquette operator in the effective Hamiltonian. 

\subsection{Geometric Factor $\alpha_p$}
The geometric factor $\alpha_{p}$ needs to be evaluated via explicit calculation and we did not find a general formula for it. 
However, we generally only care about the cases that has factor $\alpha_{p} = 0$ since it will serve as a hole in the surface code limit. We find a family of plaquettes that would have $\alpha_{p} = 0$ when the corresponding plaquette operator consists of odd number of anti-commutative check operator pairs. The proof is as follows:

Suppose we have a plaquette operator $W_p^{'}$, which is a product of several perturbation check operators.  So the factor $\alpha_{p}$ is calculated in the $l_p$th order of perturbation through the diagram(\ref{fig:perturbation}. It is easy to tell that for any permutation $\sigma \in S_{l_p}$, the $\Pi_i A_{\sigma(i)}$ has a nontrivial perturbation contribution to the plaquette operator $W_p^{'}$, where $S_n$ is $n$th order of permutation group. 

The perturbation contribution consists of two parts. One comes from the action of green function, which essentially comes from path on the perturbation tree, denoted as $Fg_{\sigma}$. Then the outcome of the perturbation could be written as $\alpha_{p} W_p^{'}=\sum_{\sigma \in S_{l_p}} Fg_{\sigma} \Pi_{i} A_{\sigma(i)}$. So $\alpha_{p} = \sum_{\sigma \in S_{l_p}} Fg_{\sigma} Sg_{\sigma}$, where $Sg_{\sigma}$ comes from making $\Pi_i A_{\sigma(i)}$ to a fixed form $W_p^{'}=\Pi^{l_p}_{i=1} A_i$, which is controlled simply by the commutation relation between $A_i$s. Consider $\bar{\sigma}: \bar{\sigma}(i)=\sigma(l_p-i+1)$, a function that is the reverse permutation of $\sigma$.  Notice that reverse the order of check operators in the perturbation tree still remains in the tree since it starts at the ground state and also ends at the ground state! And they have $Fg_{\sigma}=Fg_{\bar{\sigma}}$ due to the symmetry. 

The relative sign $\frac{Sg_{\sigma}}{Sg_{\bar{\sigma}}}$ is determined by making $\Pi_{i} A_{\sigma(i)}$ to $\Pi_{i} A_{\bar{\sigma}(i)}$. To illustrate this, look at a simple example. Consider 3 elements $A_1,A_2,A_3$, and $\sigma(i)=i$, $\bar{\sigma}(i)=3-i$. To make the product $A_1 \cdot A_2 \cdot A_3$ to $A_2 \cdot A_2 \cdot A_1$, we can swap $A_1$ with $A_2$, then $A_1$ with $A_3$, then swap $A_2$ with $A_3$. The relative sign is decided by the commutation relation of $A_i$s. The general relative sign is done by the same way. We conclude the relative sign is given by the parity of the anti-commutative pairs. The relative sign is $-1$ If $W_p$ contains odd number of anti-commutative pairs of check operators, and $+1$ if even. That results in $\alpha_{l_p}=0$ since the reverse permutation is an equivalence relation in the permutation group $S_{l_p}$.

Remark 1:
In a general lattice with vertices that $d_v=4$, we should note that the coefficient of each $W_p$ is no longer uniform. But continuously deform the coefficients of $W_p$s to be uniformly large, regardless of the sign (as we have argued, the sign does not matter regarding the toric code phase). It is a continuous transformation which does not close the gap. Therefore, it will stay in the same phase.

Remark 2:
One may worry about some unfortunate occasions that $\alpha_{p}=0$. The factor is a path-dependent factor and depends heavily on the details of the lattice and the choice of $S_c$. But we argue that, for a zero $alpha_{p}$ which have multiple channel in the diagram\ref{fig:perturbation}, we can probably find a nonzero $l_p +2$th order contribution by insert a pair of $P_{e_0}$ in the process to build up $W_p$, where $e_0 \notin Bo(p)$  into the perturbation tree, which will change the action of $G_0^{'}$ between the pair of $P_{e_0}$. Since the action is generally non-linear, it might be nonzero. If it's still zero, we could insert more. Finally $W_p$ may survive in higher order of perturbation theory.   

Remark 3: if there is huge plaquette in the lattice $\Gamma$, since $l_p$ is large, $\lambda^{l_p}$ is generally so small that this plaquette could be treated as a hole or a boundary, of which the type depends on the color of it. Or, if some $\alpha_{p}=0$, and could not survive even under the argument of remark 2, or it only survive in really high order perturbation, it also serves as a hole in the toric code model.

\section{Mapping Table of vertex with $d_v>4$} \label{generic mapping table}
Consider a vertex has degree $d_v>4$. Let us assume $d_v=2k$, so there are $k$ qubits placed on it. We call an operator a $k$-order Pauli operator if it is a tensor product of $k$ Pauli operators. We insist choosing $P_v= \bigotimes_k Z$, since the $k$ qubits would stay on the ground state of $P_v$, with eigenstate of $+1$, without loss of generality, the eigenspace would be $2^{k-1}$ dimensional. So we have to find the effective $k-1$ order Pauli operators of those $k$ order Pauli operators commuting with $P_v$. The generators of these $k$ order  are of following form:

\begin{equation}
    \mathcal{X}_i=\underbrace{1 \otimes 1 \cdots 1}_{i-1} \otimes X \otimes X \otimes 1 \otimes \cdots 1 \otimes 1
\end{equation}
Here $\mathcal{X}$ represents a tensor product of $k$ pauli operator.  $\mathcal{X}_i$ is explicitly two $X$ operator at position $i$ and $i+1$, identity $1$ at the others. We have obviously $k-1$ such operator $\{ \mathcal{X}_i|i=1,2 \ldots k-1\}$. Another set of generators are:
\begin{equation}
    \mathcal{Z}_i= \underbrace{1 \otimes 1 \cdots 1}_{i-1} \otimes Z \otimes 1 \cdots 1 \otimes 1
\end{equation}
Which means $\mathcal{Z}_i$ is one $Z$ operator locates at position $i$, Identity 1 at the others. We have $k$ such operators,
$\{ \mathcal{Z}_i|i=1,2,\ldots,k  \}$. Notice, $|\phi_0\rangle=\bigotimes_k |0\rangle$ is one of the basis in the eigenspace. $\mathcal{X_i} |\phi_0\rangle$ effectively generate the whole eigenspace. Then we define the mapping of $\mathcal{X}_i$ to $k-1$ order operator:
\begin{equation}
    \mathcal{X}_i \rightarrow \underbrace{1 \otimes 1 \cdots 1}_{i-1} \otimes X \otimes X \otimes 1 \otimes \cdots 1 \otimes 1
\end{equation}
We denote the notation $\tilde{\mathcal{X}}_i$ to represent the effective action of $\mathcal{X}_i$ in the eigenspace of $P_v$.
Notice on the right, it is a tensor product of $k-1$ Pauli operators. Similarly, $\tilde{\mathcal{Z}}_i$ to represent the effective operator of $\mathcal{Z}_i$:
\begin{equation}
    \tilde{\mathcal{Z}}_i =  \underbrace{1 \otimes 1 \cdots 1}_{i-2} \otimes Z \otimes Z \otimes 1 \otimes \cdots 1 \otimes 1
\end{equation}
For $1 < i < k-1$, and  $\tilde{\mathcal{Z}}_1 = Z \otimes 1 \cdots 1 \otimes 1$, $\tilde{\mathcal{Z}}_k = 1 \otimes 1 \cdots 1 \otimes Z $.
Notice the product of $\tilde{\mathcal{Z}}_i$ is 1, which agree with $\Pi_{i=1}^k \mathcal{Z}_i=P_v$. 
Then arranging the operators similarly as in \ref{fig:d_v=6 case} would give rise to the splitting of $d_v>6$. 

\section{Possible Measurement-Based Initializing Me-\\thod}
\label{appendix:initialization method}

\begin{figure}[htbp]
\centering
\begin{tikzpicture}[scale=0.3]
\foreach \i in {2,14,26}
\foreach \j in {3,9,15}
        \draw[thick, yellow] ({\i},{\j}) -- ({\i},{\j-3});
\foreach \i in {6,18}
\foreach \j in {3,9,15}
        \draw[thick, cyan] ({\i},{\j}) -- ({\i},{\j-3});
\foreach \i in {10,22}
\foreach \j in {3,9,15}
        \draw[thick, red] ({\i},{\j}) -- ({\i},{\j-3});
\foreach \i in {4,16,28}
\foreach \j in {6,12}
        \draw[thick, red] ({\i},{\j}) -- ({\i},{\j-3});
\foreach \i in {8,20}
\foreach \j in {6,12}
        \draw[thick, yellow] ({\i},{\j}) -- ({\i},{\j-3});
\foreach \i in {12,24}
\foreach \j in {6,12}
        \draw[thick, cyan] ({\i},{\j}) -- ({\i},{\j-3});
\foreach \i in {2,8,14,20,26}
\foreach \j in {3,6,9,12}
        \draw[thick, red] ({\i},{\j}) -- ({\i-2},{\j});
\foreach \i in {4,10,16,22,28}
\foreach \j in {3,6,9,12}
        \draw[thick, cyan] ({\i},{\j}) -- ({\i-2},{\j});
\foreach \i in {6,12,18,24,30}
\foreach \j in {3,6,9,12}
        \draw[thick, yellow] ({\i},{\j}) -- ({\i-2},{\j});

\foreach \i in {2,6,10,14,18,22,26}
\foreach \j in {3,6,9,12}
        \filldraw[black] ({\i},{\j}) circle (0.15);
\foreach \i in {4,8,12,16,20,24,28}
\foreach \j in {3,6,9,12}
        \filldraw[black] ({\i},{\j}) circle (0.15);
\foreach \i in {8,20}
\foreach \j in {1.5,7.5,13.5}
        \node[yellow] at ({\i},{\j}) {$1$};
\foreach \i in {2,14,26}
\foreach \j in {4.5,10.5}
        \node[yellow] at ({\i},{\j}) {$1$};
\foreach \i in {0,12,24}
\foreach \j in {1.5,7.5,13.5}
        \node[cyan] at ({\i},{\j}) {$2$};
\foreach \i in {6,18,30}
\foreach \j in {4.5,10.5}
        \node[cyan] at ({\i},{\j}) {$2$};
\foreach \i in {4,16,28}
\foreach \j in {1.5,7.5,13.5}
        \node[red] at ({\i},{\j}) {$3$};
\foreach \i in {10,22}
\foreach \j in {4.5,10.5}
        \node[red] at ({\i},{\j}) {$3$};
\end{tikzpicture}
\caption{This figure illustrates a honeycomb lattice used to initialize a ground state of the effective toric code, a platform that facilitates defect manipulation. Each edge is colored according to the plaquettes to which it is connected.}
\label{fig:3-colored lattice}
\end{figure}
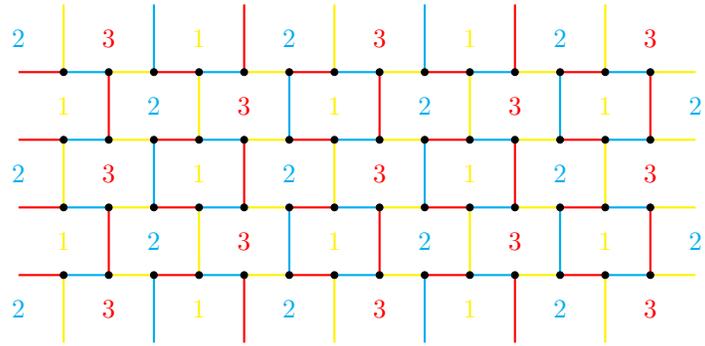

We have emphasized that since each process is described by time evolution, which is natural to depict each process (movement, creation and fusion) by quantum circuit. Here we put down a convenient measurement-based method to initialize a ground state of the effective toric code on a honeycomb lattice as in figure \ref{fig:3-colored lattice}, which is actually the same way as in the Floquet code  in \cite{Hastings_Haah_2021}. 
The method is to follow a measurement schedule:\\
Step 1: Measure the check operators associated with yellow edges.\\
Step 2: Measure the check operators associated with blue edges.\\
Step 3: Measure the check operators associated with red edges.\\
Step 4: repeat step 1.

After step 4, the state is the ground state of all plaquette terms (with signs that depend on measurement outcomes). We then proceed to measure the yellow checks (or equivalently  the elements in $S_c$ ) as depicted in fig \ref{fig:lattice without defect}. Following this measurement, the state would transition into the ground state of the plaquette operators and the stabilizer center $S_c$ (still the corresponding eigenvalues depend on the measurement outcome), thereby generating the effective toric code we want. Then we can apply the unitary operator which is given by the time evolution operators in previous section, to manipulate the twist defects.

\end{document}